\documentclass[twocolumn]{IEEEtranTCOM} 
\usepackage[dvips]{graphicx}
\graphicspath{{../eps/}{../pdf/}{../jpeg/}}
\DeclareGraphicsExtensions{.eps,.pdf,.jpeg,.png}
\usepackage[cmex10]{amsmath}
\usepackage{xcolor,amsthm, amsfonts, mdwmath, mdwtab, eqparbox, psfrag, setspace}
\usepackage[tight,footnotesize,sf,SF]{subfigure} 
\usepackage{cite,multicol, blindtext,overpic,url,color,tikz,lipsum}
\usepackage{bbm}
\definecolor{dg}{rgb}{0.1,0.55,0.15}

\usepackage{algpseudocode,algorithm}
\newtheorem{proposition}{Proposition}
\newtheorem{lemma}{Lemma}

\newtheorem{definition}{Definition}

\usepackage[pass]{geometry}

\newtheorem{remark}{Remark}
\newtheorem{corollary}{Corollary}
\usepackage{cancel,fancyhdr,acronym}
\usepackage[T1]{fontenc}
\usepackage[normalem]{ulem}
\usepackage{soul}
\usepackage{mathtools,amssymb}
\usepackage{graphicx}
\usepackage[flushleft]{threeparttable} 
\usepackage{makecell}
\usepackage{xcolor,etoolbox}
\def\@eqnnum{{\normalfont \color{red} (\theequation)}}
\makeatother

\newcommand\bout{\bgroup\markoverwith{\textcolor{blue}{\rule[.5ex]{2pt}{1.0pt}}}\ULon}
\definecolor{dg}{rgb}{0.48, 0.25, 0.0}

\usepackage{subfigure}
\usepackage{authblk}

\usepackage{xcolor}

\makeatletter
\def\ps@IEEEtitlepagestyle{
  \def\@oddfoot{\mycopyrightnotice}
  \def\@evenfoot{}
}
\def\mycopyrightnotice{
  {\footnotesize
  \begin{minipage}{\textwidth}
  \centering
  Copyright~\copyright~2017 IEEE. Personal use of this material is permitted. However, permission to use this  \\ 
  material for any other purposes must be obtained from the IEEE by sending a request to pubs-permissions@ieee.org.
  \end{minipage}
  }
}

\begin{document}
	\title{A repeated unknown game: 
		Decentralized task offloading in vehicular fog computing
		\author{\IEEEauthorblockN{Byungjin Cho and Yu Xiao}	\vspace*{-.8cm}	
			\thanks{		
				This work has received funding from Academy of Finland under grant number 317432 and 318937.
				
				Byungjin Cho and Yu Xiao are with the Department of Communications and Networking, Aalto University, 00076 Espoo, Finland (e-mail: byungjin.cho@aalto.fi; yu.xiao@aalto.fi).	
			}
		}
	}
	\maketitle

	\begin{abstract} 
		Offloading computation to nearby edge/fog computing nodes, including the ones carried by moving vehicles, e.g., vehicular fog nodes (VFN), has proved to be a promising approach for enabling low-latency and compute-intensive mobility applications, such as cooperative and autonomous driving. This work considers	vehicular fog computing scenarios where the clients of computation offloading services try to minimize their own costs while deciding which VFNs to offload their tasks. 
		We focus on decentralized multi-agent decision-making in a repeated unknown game where each agent, e.g., service client, can observe only its own action and realized cost. In other words, each agent is unaware of the game composition or even the existence of opponents. We apply a completely uncoupled learning rule to generalize the decentralized decision-making algorithm presented in \cite{Cho2021} for the multi-agent case. The multi-agent solution proposed in this work can capture the unknown offloading cost variations susceptive to resource congestion under an adversarial framework where each agent may take implicit cost estimation and suitable resource choice adapting to the dynamics associated with volatile supply and demand. According to the evaluation via simulation, this work reveals that such individual perturbations for robustness to uncertainty and adaptation to dynamicity ensure a certain level of optimality in terms of social welfare, e.g., converging the actual sequence of play with unknown and asymmetric attributes and lowering the correspondent cost in social welfare due to the self-interested behaviors of agents.
		 
	\end{abstract} 

	\vspace{-.35cm}	
	\section{{Introduction}}
	
	{
		Emerging vehicular applications, such as autonomous and cooperative driving, require low-latency networking and computing services. Besides cellular vehicle-to-everything (C-V2X) \cite{3gpp2018}, edge/fog computing is another key enabling technology that brings computing resources close to end users like connected vehicles \cite{Mao2017, Mouradian2018}. End users with relatively low computing power can offload compute-intensive tasks to nearby edge/fog computing nodes. In practice, edge/fog computing nodes can be installed in radio access networks, or on vehicles. The latter is called vehicular fog nodes (VFNs) in this paper, utilized as a viable component of a new computing paradigm, vehicular fog computing (VFC)\cite{Hou2016,Zhu2018}. {Due to the mobility of both end users and VFNs, the network topology regarding communications between VFNs and end users is unstable \cite{Chen2019, Cho2021}. This poses challenges to task offloading decision making, i.e., where to offload tasks.}

		Task offloading decision making can be implemented in either a centralized or a decentralized manner \cite{Zheng2015, Gu2018, Zhou2021a, Jehangiri2022, Feng2018}. In the first case, a controller/coordinator collects state information from all the fog nodes and decides which tasks should be offloaded to which fog nodes, with the aim of minimizing the service costs, using a stochastic control process~\cite{Zheng2015, Gu2018} or non-probabilistic approach~\cite{Zhou2021a, Jehangiri2022}. In the latter, each task generator, i.e., a service client that generates computational tasks to be offloaded, independently decides where to offload its tasks based on its local observation~\cite{Feng2018}. Compared with the centralized approach which causes extra signaling overhead and raises privacy concerns, the decentralized approach has been considered a more promising solution, given the condition that it can dynamically adapt to the changes in the environment, e.g., network topology, the availability of computing capacity, and the variation in demand.  
		 	
		Regarding the decentralized approach, many efforts have been invested in enabling task generators to learn directly the states of VFNs and to make offloading decisions based on history \cite{sun2019,Zhu2019, Liu2020, Zhang2020, Cho2021}, i.e., clients learn as much as possible about different candidate actions that lead to good estimates of their costs, and simultaneously optimize the desired objective to select the optimal actions given the learned information. In such learning processes, there are two fundamental issues to address. One is to balance the self-exploration/exploitation trade-off in the learning process for improving its efficiency, i.e., making decisions with the aim of reducing uncertainty over states. The other is to validate whether individual-level learning dynamics of distributed task generators would lead to a system-level equilibrium in some sense \cite{Xiao2020}. Such no-regret decisions made by self-interested individuals generally conflict with collective desires at the system level.  
		
		 Game theory, designed for strategic interactions among decision making agents sharing scarce resources, e.g., service clients competing for limited VFN resources, has been used to analyze individual-collective conflicts and to formulate the alternative strategy to compete with one another \cite{sun2021}. Previous works along this line have mainly focused on characterizing game-theoretic equilibria and their efficiency, and deriving distributed learning algorithms that converge to the desired outcomes. Most of the existing results, however, are based on the following two assumptions: i) each agent has complete information about the composition of the game-theoretic settings, i.e., actions and costs of the other service clients are observable and known, and ii) one may always face the same game repeated over time, e.g., symmetric game structure. While these assumptions lead to strong theoretical guarantees, e.g., converging towards the stable and efficient equilibrium, it is often unrealistic since decentralized task offloading systems in vehicular environments are inevitably limited in terms of access to and use of such system-level knowledge, and are inherently dynamic and heterogeneous. Thus, individual agents with different resource preferences and availability due to volatile VFNs, may not obtain system-level optimality.
		 
		Motivated by the concerns above, this work formulates the decentralized multi-agent decision-making in terms of VFN selection, as a repeated unknown game where each agent i) has access to only local information such as its own actions and utilities, but is unaware of the game composition or even the existence of opponents, and ii) adapts its offloading decision to non-stationary and arbitrary dynamics. In the literature, such unknown and asymmetric attributes have not been considered due to \textit{the challenges associated with characterizing games that lack convergence to equilibrium and its efficiency}. We overcome the challenges by embracing a learning-based decentralized offloading algorithm based on a variant of adversarial multi-armed bandit (MAB) for a general multi-agent scenario with regard to the unsung game dynamic. Our solution is able to capture the unexplored offloading cost variations, adapt to the evolving circumstance, and achieve a better balance between individual-level acceptability and system-level efficiency. The main contributions of this work are summarized below.
		 		
		\begin{itemize}
			\item This work extends the study presented in \cite{Cho2021} where a single agent competes with a black-box adversarial environment for self-interested regret-optimal, to a multi-agent case where dynamical behaviors of distributed self-interested individuals contending with other anonymous agents and adversaries desire to achieve a certain level of optimality in terms of social welfare.
						
			\item This work allows each agent to provoke adjustment dynamics in a manner adaptive to volatile resource necessitate and provision, e.g., variations in requested workload and candidate VFN set. Such independent amendment to dynamicity may threaten community stability due to aggravated uncertainty. We have conducted a convergence analysis of decentralized strategies for the offloading game, the actual sequence of play. 
			
                \item This work maps the decentralized learning dynamics of individual agents to the unsung system-level equilibria with asymmetric properties: all agents may have i) unequal learning rates, ii) asynchronous update times, iii) dissimilar candidate VFNs, and iv) uneven implicit explorations. The actual sequences of strategies induced by the asymmetric explorations have proved to converge to a small neighborhood of the equilibrium point.		 
						
			\item This work shows that individual perturbations for robustness to uncertainty (implicit exploration) and adaptation to dynamicity ensure certain system-level optimality, i.e., while reaching a sequence of stable equilibrium points, the self-interested behaviors inclining toward robustness and adaptation allow for lowering the upper bounds of the price of total anarchy (PoTA). Extensive simulation results verify its effectiveness. 			
			
	\end{itemize}}
	The rest of this paper is organized as follows: 
	Section II presents the background and related works. Section III describes a system and problem formulation. Sections IV and V show offloading strategy and convergence analysis. Sections VI and VII include simulation results and conclusions.

	\section{{Background and Related work}}
 	In this section, we introduce the necessary game-theoretic background
	and review related works. The summary of the parameter symbols is given in Table~\ref{tab:symbols}.
	 	
	\begin{table*}[t]
		\centering
		\caption{List of parameters}
			\begin{threeparttable}
				\centering
				\resizebox{.975\textwidth}{!}
				{%
					\begin{tabular}{|c|l|} \hline
						Parameter & Description \\ \hline  
						$\mathcal{T}, \tau$ & Set $\mathcal{T} = \{1,\cdots,\tau,\cdots,|\mathcal{T}|\}$ where $\tau$ denotes $\tau$-th time frame
						\\ \hline  
						$\mathcal{N}, n$ & Set $\mathcal{N} = \{1,\cdots,n,\cdots,|\mathcal{N}|\}$ where $n$  denotes $n$-th client     
						\\ \hline  
						$\mathcal{K}, k$ & Set $\mathcal{K} = \bigcup_n \mathcal{K}_n  = \{1,\cdots,k,\cdots,|\mathcal{K}|\}$, where $\mathcal{K}_n$ is a subset of $\mathcal{K}$, VFNs of $n$-th client, and $k$ denotes $k$-th VFN \\ \hline  
						$ {l}$, {${l}_{n}$, ${l}_{nk}$} & Vectors $ {l}=[l_n]_{n\in\mathcal{N}}$ where $l_n = [l_{nk}]_{k\in\mathcal{K}}$ is a cost vector of $n$-th client, {and $l_{nk}$ is a unit cost for $n$-th client offloading to $k$-th VFN} \\ \hline  
						$R_n$ & Regret of $n$-th client\\ \hline
						$F_k$, $f_{nk}$, $c_k$ & Maximum CPU frequency of $k$-th VFN, its allocation to $n$-th client of $k$-th VFN, and its congestion degree\\ \hline
						$q_n$, $w_n$ & Input data size of $n$-th client, and its computation complexity     \\ \hline
						$ \delta_n$, $\zeta_n$ &  Per-task demand measure of $n$-th client, and its normalized weight with per-bit demand measure\\ \hline
						$\beta_{nk}$, $\tau_o$ &	Patched learning score of a arm $k$ newly appearing at $\tau_o$	for $n$-th client \\ \hline
						$\eta_n'$, $\gamma_n'$, $\vartheta_n$, $\kappa$ &	
						Learning and implicit exploration rates of $n$-th client, the number of times a client $n$ has been involved in an interaction, reference rate\\ \hline
						$\hat{l}_{nk}$, $\hat{\mathcal{L}}_{nk}$, $\mathcal{W}_{nk}$ &	
						Estimated cost of $l_{nk}$, weighted learning score, cumulative weighted score of an arm for $n$-th client\\ \hline
						${p}$, {${p}_{n}$, ${p}_{nk}$} & Vectors $ {p}=[p_n]_{n\in\mathcal{N}}$ where $p_n = [p_{nk}]_{k\in\mathcal{K}}$ is a probability vector of $n$-th client, {and $p_{nk}$ is a probability that $n$-th client selects $k$-th VFN}  \\ \hline  
						$ {p}_{nk}(\tau), \dot{{p}}_{nk}$ & Discrete and continuous time processes for $p_{nk}$, respectively. \\ \hline  
						 						
					\end{tabular}%
				}
			\end{threeparttable}	
			\label{tab:symbols}
		\end{table*}

		\begin{table*}[t]
			\centering
			\caption{Comparison of offloading strategies for target properties (stability, scalability, personality, adaptivity)}
				\begin{threeparttable}
					\centering
					\resizebox{.980\textwidth}{!}{%
						\begin{tabular}{|c|c||c|c|c|} \hline
							{\bf Works}       
							& {\bf \begin{tabular}[c]{@{}c@{}} System performance (stability)\\ actual play/efficiency \end{tabular}}
							& {\bf \begin{tabular}[c]{@{}c@{}}Uncerntainty (scalability)\\  noncooperative/bandit-feedback/non-stochastic\end{tabular}} 			
							& {\bf \begin{tabular}[c]{@{}c@{}}Heterogenity (personality)\\ learning rate/update/estimation/action \end{tabular}}  
							& {\bf \begin{tabular}[c]{@{}c@{}} Dynamicity (adaptivity) \\ resource supply/demand\end{tabular}} 
							\\ \hline \hline
							\cite{Chen2016}              & $\bigcirc$   & $\bigtriangleup$           & $\times$    & $\times$     \\ \hline
							\cite{Zheng2018}             & $\bigcirc$   & $\times$           & $\bigtriangleup$    & $\times$     \\ \hline
							\cite{Josilo2019}             & $\bigcirc$   & $\bigtriangleup$           & $\bigtriangleup$    & $\times$     \\ \hline 
							\cite{Sergiu2000}                    &$\bigtriangleup$          & $\bigtriangleup$  & $\times$ & $\times$   \\ \hline    
							\cite{Kleinberg2009, Krichene2015, Palaiopanos2017}  & $\bigtriangleup$ & $\bigcirc$          &  $\bigtriangleup$ &  $\times$   \\ \hline
							\cite{Coucheney2015, Cohen2017}           &$\bigtriangleup$  & $\bigcirc$  & $\times$  &  $\times$     \\ \hline
							\cite{Milchtaich1996, Feng2017} &$\bigtriangleup$   &$\bigtriangleup$  & $\bigtriangleup$   & $\times$  \\ \hline  
							\cite{Gummadi2013}  & $\bigtriangleup$   & $\bigtriangleup$        &  $\times$  &  $\times$       \\ \hline
							\cite{Shi2021}    &  $\bigtriangleup$       &$\bigtriangleup$  & $\times$   & $\times$ \\ \hline 
							\cite{Zhou2021}    &  $\bigtriangleup$       &$\bigcirc$  & $\times$   & $\bigtriangleup$ \\ \hline 
							\cite{Cho2021}     & $\times$   & $\bigcirc$    & $\times$   & $\bigcirc$        \\ \hline

							\bf This work      & $\bigcirc$   & $\bigcirc$    & $\bigcirc$   & $\bigcirc$   \\ \hline
						\end{tabular}%
					}
					\begin{tablenotes}
						\item[*] \footnotesize{The symbols, $\bigcirc$, $\bigtriangleup$, and $\times$, represent that considered aspects for each target property are fully, partially, and minimally addressed, respectively.}
					\end{tablenotes}
				\end{threeparttable}	
				\label{tab:related_work_comparison}
			\end{table*}

			\subsection{Background: games and dynamics}\label{background:background}
			\subsubsection{Games}
			An $|\mathcal{N}|$-agent (task offloading) game is represented by a tuple		$<\mathcal{N}, \mathcal{K}, {l}>$ where $\mathcal{K}$ is a set of actions (candidate clients) and a set of agents (clients) $\mathcal{N}$ decide individually which action (VFN) to choose from $\mathcal{K}$. ${l} = [l_1, \cdots, l_{|\mathcal{N}|}]$ refers to payoff ({unit cost of task offloading}) vectors of agents in $\mathcal{N}$, while $l_{nk}$ is the payoff for agent $n$ to take the action $k$ ({unit cost of offloading a task from agent $n$ to an action $k$}). The preferences of the $n$-th agent for one action over another are determined by an associated payoff function $l_n$: $\mathcal{K} = \prod_{n} \mathcal{K}_n \rightarrow \mathcal{R}$ that maps the action profile of all agents' chosen actions to the agent's payoff. Every agent can mix its actions by playing probability distributions over its action sets. A mixed strategy of agent $n$ is a probability vector, $p_n = [p_{n1}, \cdots, p_{n|\mathcal{K}|}] \in \mathcal{X}_{n}$ where $\mathcal{X}_n$ is the mixed strategy space of agent $n$. The space of all mixed strategies over agents induces a strategy profile $p$, ${p} = [p_1, \cdots, p_{|\mathcal{N}|}]\in \mathcal{X} = \prod_{n\in\mathcal{N}} \mathcal{X}_n$. 	
			\begin{definition} \label{def_ne}
				Agents in a game may reach a state where no one can further reduce its own cost by changing strategies. If {$l_n({p}_{n}'; p_{-n}') \!\leq\! l_n({p}_n; {p}_{-n}')$}\footnote{$(p_n; p_{-n})$ stands for ${p}\in \mathcal{X}$ used to highlight the strategy of agent $n$ against that of all other agents.} for every deviation $p_n \in \mathcal{X}_n, \forall n \in \mathcal{N}$, a strategy profile ${p}'\in \mathcal{X}$ is a Nash equilibrium (NE).
			\end{definition}

			In a game played over $\mathcal{T}$ rounds, after taking an action $k_n$ potentially randomized according to a mixed strategy in the $\tau\in\mathcal{T}$ round, agent $n$ receives a payoff and may observe the payoff vector for all actions in its action space against the selected actions of the other agents. Each agent may be unwilling to disclose their private information, and it independently selects an action for a given task without cooperation with each other, i.e., the actions and resulting costs of the other agents are unknown and unobservable. 
			
			\begin{definition} \label{def_unknown}
				(Repeated unknown game) In a repeated unknown game, each agent does not know the payoff function of any agent (including itself), and after each round, each agent receives its own costs but it sees neither the choices of the other agents nor the resulting payoffs.  
			\end{definition}
			 
				The social welfare of an outcome is defined as the sum of the individual payoffs of the agents. The price of anarchy (PoA) is the well-known metric used in game theory~\cite{Roughgarden2015}, to quantify the inefficiency of the selfish behavior of the agents in such an information-limited situation, that is, how bad the social welfare at equilibrium is as compared to the optimum social welfare achievable. It is defined as the worst possible ratio between the social welfare of a NE and  that of any optimal strategy, expressed as $ {{C}}/{C^*}$ where $C =  \mathbb{E}_{{k'} \sim {{p}}}[\sum_{n\in\mathcal{N}} {l}_{nk'}]$ is the expected social utility over the randomness of the clients, and $C^*$ is the value of socially optimum strategy profile, ${{p}}^* = \arg \inf_{{p}\in \mathcal{X}} C$. In order to analyze the overall efficiency loss in repeated play, a generalization of PoA, called the price of total
				anarchy (PoTA), is used \cite{Xiao2020}, defined as the ratio between the average total cost over a period of time and the total cost of the optimal outcome, expressed as $\mbox{PoTA} =\frac{1}{|\mathcal{T}|} \sum_{t\in \mathcal{T}} \sum_{n\in\mathcal{N}} l_{nk'}(\tau) / C^*$.

			\subsubsection{Dynamics}
			From a single agent's perspective, a repeated unknown game can be viewed as an online learning problem in which the agent selects actions sequentially by learning from past experiences. 
			Note that online learning under uncertainty relies on feedback in general. Thus, the quality of the feedback in terms of completeness has significant implications in learning rule. In the case of full-information feedback, payoffs (costs) of all actions (VFNs) an agent could have taken are observed in each stage. Incompleteness could be temporal across decision states and it could be also spatial across the action space. One popular model of incomplete feedback is the so-called bandit feedback, where only the payoff of the chosen action is revealed. The term bandit feedback has its roots in the classical single agent online learning problem of playing a multi-armed slot machine known as a bandit (action). 
			An agent determines which arm to pull for each time frame, but faces the exploration and exploitation trade-off. The agent has to decide between exploring actions to obtain information about the environment and selecting the action that has historically given the lowest cost. The individual performance of a client $n$ is quantified via regret by measuring the cumulative cost against a benchmark policy, e.g., the learning objective that an online algorithm aims to achieve over time. In case of the best policy in hindsight, regret is a metric that measures the differences of the costs from the agent to those from the best action, $R_n(|\mathcal{T}|) = \mathbb{E}\left[\sum_{\tau\in \mathcal{T}} l_{nk'}(\tau)\right]- \mathbb{E}\left[\sum_{\tau\in \mathcal{T}} l_{nk''}(\tau)\right]$ where $k'$ is the solution obtained by an online learning algorithm and $k''$ is the optimal solutions given that $ k''= \arg\min_{k\in\mathcal{K}_n(\tau)} \mathbb{E}\left[\sum_{\tau\in \mathcal{T}} l_{nk}(\tau)\right], \forall n$, for the sequence of time frames $\mathcal{T}$ during which the actions are identical \cite{sun2019}.

			\subsection{{Related work}}
			
			We present the previous works related to offloading algorithms using a game theoretic approach and online learning algorithms for multi-agents, summarized in  Table~\ref{tab:related_work_comparison}.
			
			Several works have tried to solve the decentralized task offloading problem in a game-theoretic context. For example, the work in \cite{Chen2016} formulated the decision-making process as a potential game where a distributed algorithm was designed to decide computation offloading and select a proper resource, e.g., wireless channel. The work in \cite{Zheng2018} utilized a stochastic game to decide the actions of offloading clients in a distributed manner. The work in \cite{Josilo2019} proposed a Stackelberg game that considered the interaction between a central controller (leader) and clients (followers). While the clients act as decentralized decision makers on whether or not to use shared resources to offload, the central entity coordinates the offloading decisions of the clients such that the resources are efficiently utilized. However, the aforementioned research has been built on impractical assumptions listed below. 	
			Firstly, they assumed that the environmental state could be obtained by exchanging signaling information between resource requesters and providers \cite{Chen2016,Zheng2018,Josilo2019}, or forecasted in a stochastic domain \cite{Josilo2019}, which is not suitable for the case where the immediate environmental dynamics and action state relevant parameters are not available. 
			Secondly, they ignored \textit{the personal, self-interested properties in terms of exploration strategies}, e.g., asymmetric update dynamics with different candidate action sets.			
			Thirdly, they overlook \textit{the individual perturbations for improving self-objectives} adapting to resource demand and supply dynamics and their impacts on the system-level performance, e.g., in terms of convergence to a NE and its effectiveness.

			Online learning strategy in a multi-agent framework has both similarities and differences from the single agent one. When one agent is isolated by abstracting away all the other agents, an online learning problem of a learner could be re-established. In this case, the agent's regret is most commonly used as a quality indicator \cite{Cho2021}. From a game-theoretic standpoint, however, the main question that arises is whether agents eventually settle on an equilibrium profile from which no agent has the incentive to deviate. It has been known that if every agent adopts a no-regret learning algorithm, the sequences of actions taken by all agents converge to the weak or restrictive set of correlated equilibria \cite{Sergiu2000}. The convergence outcomes are the empirical frequency of the agents' actions. From a game-theoretic perspective, even if such time-averaging results might still converge, the actual sequence of play may fail to converge altogether, so the agents' actual behavior and the payoffs they obtain could be different. Thus, it is crucial to establish \textit{convergence of the actual sequence of play generated by an online learning process, rather than leveraging the time-average approach, particularly in a dynamic environment}.

			The convergence of actual sequences has been studied with variant algorithms for no-regret learning. The works in \cite{Kleinberg2009, Krichene2015, Palaiopanos2017} showed that agents end up playing an equilibrium with probability arbitrarily close to 1, and the actual sequences induced by the multiplicative weights update rule converges to an equilibrium with a state-dependent diminishing learning step \cite{Kleinberg2009, Krichene2015}, or with only a small enough constant step-size \cite{Palaiopanos2017}. While, in the above works, agents are assumed to have full knowledge of their payoff vectors, including actions that were not chosen, variants of the exponential weights algorithm in a minimal-information setting, e.g., bandit feedback, converge to a relaxed NE, based on unbounded or truncated estimators \cite{Coucheney2015, Cohen2017}. However, all existing works are valid in the context of potential games admitting the finite improvement property (FIP) with which every single-agent improvement path sequence decreases the potential by the same amount as the agent's cost, and terminates at a NE in finite steps. \textit{A general unknown offloading game with asymmetric and client-specific features}, which does not generally possess the FIP \cite{Feng2017} except for special cases, e.g., with only  2 agents or 2 VFNs \cite{Milchtaich1996}, is considered in this work.
			
			{Existing works \cite{Gummadi2013} on multi-agent MAB in a repeated game achieve system stability based on the assumptions of state regeneration and stochastic settings. While the state regeneration processes account for the situation where some system entities join or leave the game, too frequent re-setting in a dynamic environment may invalidate the learning process. A well-behaved stochastic model exists for each arm, but it is often difficult to determine the correct stochastic assumptions in real-world applications. Different non-stochastic MAB settings are explored in \cite{Shi2021, Cho2021, Zhou2021}. Nevertheless, these works i) leverage information exchange with other agents (lack of scalability) \cite{Shi2021}, ii) overlook heterogeneous individuality with an assumption of the fixed learning rate, synchronized update time, and explicit exploration-based estimation (lack of personality) \cite{Zhou2021}, iii) do not guarantee the system-level performance in presence of heterogeneous learning dynamics (lack of system-level performance) \cite{Cho2021}, and iv) commonly do not address the stability analysis with \textit{individual perturbations resistant to uncertainty and adaptive to the agent's demand and arm's availability}.}

			\section{System model and problem formulation}
			This section presents the system model of the task offloading scenario and the formulation of the offloading problem.

			\subsection{System model}
			  
			\begin{figure}[t]
				\centering \hspace{-0.05em}
				\includegraphics[width=0.2037022\textwidth]{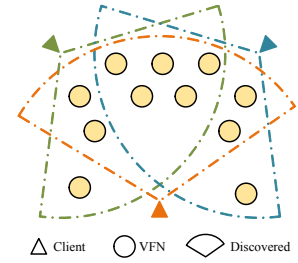}
				\caption{System model with multiple service clients and candidate VFNs. As an example, among 3 clients, one has access to 9 VFNs, while the others have 8 VFNs, $|\mathcal{N}| = 3$, $|\mathcal{K}| = 11$, $|\mathcal{K}_n| = 9$, and $|\mathcal{K}_{-n}| = 8$.}\vspace{-0.9em}
				\label{scenario}
				\end{figure}
			
			Consider a system shown in Fig. \ref{scenario}. A set of service clients $n \in \mathcal{N} = \{1,...,|\mathcal{N}|\}$ generate computation tasks which are meant to be offloaded, while a set of VFNs (fog nodes), $k \in \mathcal{K} = \{1,...,|\mathcal{K}|\}$ are candidates for handling the offloaded computational tasks. Assume that at time $t$, a client $n$ can offload a task to a VFN, $k \in \mathcal{K}_n({t}) \subseteq \mathcal{K}$, within the client's communication range. Due to inherent mobility, the set of candidate VFNs $\mathcal{K}_n({t})$ varies over time. 
			The candidate sets in the next time slots are unknown prior due to unknown mobility behaviors. We assume that $\mathcal{K}_n({t})\neq \emptyset, \forall n, t$. Otherwise, the task may be processed locally or forwarded to a remote cloud server. Each client forms the candidate set based on the topological states such as speeds, locations, and moving directions of the client and the VFNs \cite{Zhu2018}. Such status information can be acquired by other neighboring vehicles through a vehicular communication protocol, e.g., C-V2X, while it does not include the availability of computing capacity. To better characterize the client and fog nodes' movement, without loss of generality, we assume that the VFN selection for task offloading is scheduled periodically. The timeline is discretized into time frames $\tau\!\in\! \mathcal{T}\!\! =\!\! \{1,..., |\mathcal{T}|\}$, and each client can select at most one VFN in each time frame if it has a task to offload.

			In general, computational tasks can be partitioned into sub-ones at different granularity levels\cite{2016Akherfi}. In this work, each task is considered as a basic unit for offloading, i.e., offloaded to and processed by a VFN within one-time frame $\tau\in[\tau, \tau+1)$ \cite{Zhang2020}. One can characterize a task for $\tau$ with two parameters, including the input data size $q_n({\tau})$ (bits/task) and the number of CPU cycles required to process one-bit input for $\tau$, $w_{n}({\tau})$ (cycles/bit). The input size is bounded by client-specific limits, e.g., $\underline{q}_{n}\leq q_{n}({\tau})\leq \bar{q}_{n}$ where $\underline{q}_{n}$ and $\bar{q}_{n}$ are the lower and upper thresholds on the input data size of a client $n$, respectively. The value of $w_{n}({\tau})$ varies depending on the nature of performed applications. Assuming that a similar type of application is used for all clients, one may have $\bar{q}_{n}= \bar{q}$ and $\underline{q}_{n}= \underline{q}$, with individually different demand, $\underline{q}\leq q_{n} (\tau)\leq \bar{q}$.

			The computational capacity of a VFN $k\in \mathcal{K}_n({\tau})$ is measured by its maximum CPU frequency $F_k$ (cycles/second). One task is offloaded as a whole to a VFN. Each VFN may execute tasks in parallel depending on its own resource allocation rules unknown to clients. It may also adjust its CPU frequency in a dynamic manner, e.g., with dynamic frequency and voltage scaling technique. This work considers that the computing capacity of a VFN $k$ allocated to a client $n$ for $\tau$, denoted by $f_{nk}({\tau})$ (cycles/second), is determined by the computing resource allocation policy, and remains static for each $\tau$. The available computing resources of a VFN $k$ is non-increasing w.r.t the total number of clients offloading to the same VFN $k$ \cite{sun2019, Gummadi2013}. It can be also arbitrarily constrained by an attacker, i.e., a malicious adversary may inject fake tasks into a VFN and disturb this allowable resource allocation. 
           The wireless resource is orthogonally assigned to each task. The achievable uplink and downlink transmission rates between a client $n$ and a VFN $k$ are affected by the physical characteristics such as the distance between a VFN and a client, and the amount of allocated bandwidth.

		\subsection{Problem formulation}
		Performing task offloading incurs transmission and computation costs. For task offloading to a fog node, the end-to-end latency of the task originates from a linear combination of the following segments: i) generation of the computational tasks, ii) uplink transmission latency, iii) processing latency at the fog node, iv) downlink transmission latency and v) processing of the results received from the fog node. It is often assumed that the size of the computational results is small enough that the downlink transmission latency can be safely ignored \cite{Zhang2020}. We define the unit cost of offloading a task from client $n$ to a VFN $k$ at time frame $\tau$ as $l_{nk}(\tau)$. It calculates the overall delay caused by transmitting one bit of input data to $k$ and processing it on $k$. The processing delay is calculated as the number of CPU cycles required divided by the CPU frequency of $k$.  The unit cost of task allocation $l_{nk}(\tau)$ can be written as	
		\begin{equation} \label{eq:P00}
			l_{nk}({\tau}) = {1}/{r_{nk}({\tau})} + {w_n(\tau)}/{f_{nk}(\tau)}, 
		\end{equation}
		where $r_{nk}(\tau)  = B\log\left[1+ \frac{P g_{nk}}{N + I_k}\right]$ is the link rate for  transmitting a task from a client $n$ to a VFN $k$ {at the time frame $\tau$}, $B$ is channel bandwidth, $P$ denotes the transmission power of $n$, $g_{nk}$ is the uplink channel gains between $n$ and $k$, $N$ is the noise power, and $I_k$ denotes interference measured at $k$. Given the orthogonal channel allocation \cite{Kenney2011}, the co-channel interference can be avoided. Furthermore, the cross-channel interference can be ignored according to the experimental results in \cite{Rai2007}. The channel gains are static during the uploading process of each task. The unit cost is dominated by the computation part in particular for high-computational complexity tasks \cite{Zhang2020}, e.g., high $w_n (\tau)$.
 
		We aim at minimizing the expected unit cost of task allocation across finite time frames by guiding clients to make individual decisions about where to offload a task in each time frame. The workload of fog nodes in dynamic and heterogeneous networks is hard to predict, and exchanging such information among the vehicles and fog nodes causes high signaling overhead. Thus, the client may lack the state information of fog nodes and could not make an accurate estimate about which fog node would provide the optimal offloading service. To overcome this, one may utilize a learning-and-adapting-based offloading scheme where a client observes and learns the performance of candidate fog nodes and makes an offloading decision based on historical observations without exact knowledge of the current state information. For this, each client $n$ makes use of a learning-based algorithm \cite{Cho2021} focusing on task offloading problem to minimize the expected cumulative unit cost of all tasks, expressed as
		\begin{eqnarray} \label{eq:P0}
			\mathcal{P}: \min_{\boldsymbol{k}'} \mathbb{E}\left[\sum_{{\tau\in \mathcal{T}}} l_{n {k}'}({\tau})\right],
		\end{eqnarray}
		where $\mathbb{E}\left[\cdot\right]$ is the expectation operator, $l_{nk'}({\tau})$ is a sequence of unit costs for $\forall \tau\in \mathcal{T}$ {in equation (\ref{eq:P00})}, and $\boldsymbol{k}'$ is a sequence of optimization variables where each variable represents the index of a fog node selected at a time frame $\tau$, $k'\in \mathcal{K}_n({\tau})$. The index $\tau$ is omitted below for ease of description.

		
        \section{Learning based offloading strategies in a repeated unknown game}
        The VFN selection problems of multiple service clients can be modeled as a repeated unknown game, $\Gamma =<\mathcal{N}, \mathcal{K}, {l}>$. This section presents learning dynamics in the game $\Gamma$, from a multi-agent and an individual amendment points of view.
			
		\subsection{Completely uncoupled learning dynamics for multi-agents}
			Multi-agents in a repeated unknown game achieve their own individual and societal values through their learning dynamics. 
			
			\subsubsection{Adversarial MAB for single agent} 
			A task requester performs an online learning process while running the offloading service and updating the optimal decision on the action (VFN) selection. The objective of a single agent (client) $n$ is to minimize the long-term cost as shown in equation (\ref{eq:P0}), while managing the exploration-exploitation trade-off. Such an exploration versus exploitation dilemma can be formulated as a MAB problem where each neighboring VFN is treated as an independent arm, and each arm generates cost in an adversarial fashion. Upon making the decision, the agent $n$ receives a realized cost chosen by an adversary.
   
			Unlike a classical stochastic MAB whereby the costs are generated randomly and independently following a fixed but unknown distribution, no statistical assumption about the generation of the cost is required in an adversarial MAB. In adversarial MAB problems, a randomized selection policy is needed due to the possibility that an agent using a deterministic policy may be easily fooled by adversaries. Thus, instead of choosing an action $k \in \mathcal{K}_n(\tau)$ directly, the agent selects a probability distribution over the available actions for $\tau$, $p_n (\tau)\in [0,1]^{|\mathcal{K}_n(\tau)|}: \sum_{k\in \mathcal{K}_n(\tau)} p_{nk}(\tau) = 1$ and draws an action $k'$ according to this distribution, $k' \sim p_n$. The selected probability distribution is proportional to its loss, weighted appropriately. The intuition is to give more weight to actions that performed well in the past. Weighted-average randomized strategies with potentials could be considered to achieve a cumulative cost (almost) as small as that of the best action \cite[Section 6]{cesa2006}. An action $k$ is assigned with the selected probability for task $\tau$, $p_{nk}(\tau)$ which is proportional to weighted accumulated cost caused by that action in the past, $p_n(\tau) = \left[\frac{W_{nk}(\tau)}{\sum_{m} W_{nm}(\tau)}\right]_{\forall k}$, where $W_{nk}(\tau)$ denotes a weight of each action $k$ maintained by the agent $n$, representing the confidence that $k$ is a good choice for $n$. 
			In a bandit setting, rather than concerns about how to get the estimated cost of an action (arm) that was not pulled, one seeks to investigate how such information can be used when it becomes available. To that end, the score-based learning process is considered as follows: {The service capacity of an action can be represented by the score parameter, the cumulative estimated per-bit cost up to $\tau$, $\mathcal{\hat{L}}_{nk}({\tau}) = \sum_{\tau'=1}^{\tau}   \eta_n'({\tau'}) \hat{l}_{nk}({\tau'})$, where $\eta_{n}'({\tau'})\in (0,1]$ is the learning rate of agent $n$ for $\tau'$, and $\hat{l}_{nk}({\tau'})$ is the estimated cost from the action $k$ for $\tau'$.} Considering the exponential potential function with the score, the logit choice map $\Lambda_n: \mathcal{R}^{|\mathcal{K}_n|} \rightarrow \mathcal{X}_n$ yields a probability vector,  $p_{n}(\tau) = \left[\frac{{W}_{nk}(\tau)}{ \sum_{m\in \mathcal{K}_n(\tau)} {W}_{nm}(\tau)}\right]_{k\in \mathcal{K}_n(\tau)}$, where ${W}_{nk}(\tau) = e^{-\mathcal{\hat{L}}_{nk}(\tau-1)}$. The scores reinforce the success of each strategy measured by the estimated cost $\hat{l}_{nk}(\tau-1)$, so an agent would rely on the strategy with the lowest score.

		\subsubsection{Adversarial MAB for multi-agents} 
		We generalize the single agent adversarial MAB by considering a set of agents $n\in\mathcal{N}$. Note that the realized cost for an arm, a VFN in case of offloading decision making, is affected by an adversary and congestion effects \cite{Shi2021}. A third entity (or even another agent leveraging the same arm) could be an adversary rendering an arbitrary sequence of service costs. A VFN's computational capacity allocated to an agent $n$ depends on the number of other agents offloading their tasks to the same VFN and the computing resource allocation policy in use. 
			The cost sequence forced by the adversary could be independent or dependent on the actions of agents, e.g., oblivious or non-oblivious.  
			The realized costs for different agents choosing the same action might be different due to other attributes such as the dynamic computing resource allocation rules and different communication costs. Such unignorable effects caused by the asymmetric factors make the oblivious adversary approach in an unknown game reasonable.   
			In general, the cost of selecting an arm $k$ is non-decreasing w.r.t its congestion degree. The congestion degree of an arm $k$ is defined as $c_k = 1+\sum_{u\in\mathcal{N}\backslash n} \mathbbm{1}(k = k_u)$, where $k\in\mathcal{K}_n$ and $k_{u}\in\mathcal{K}_{u}$ are the arms chosen by an agent $n$ and the other agents $u\in\mathcal{N}\backslash n$, respectively. Existing works in \cite{Rosenski2016, Shi2021} take a common assumption that any collision\footnote{It occurs when more than one agent pulls the same arm simultaneously, e.g., $c_k>1$. Any collision with another agent is assumed to be perfectly known in the literature.} with other agents could be avoided/resolved through communication between agents or collision indicator-like information $\mathbbm{1}_{c_k}= \mathbbm{1}(c_k>1)$, e.g., a binary value of the realized loss presumably indicating collision occurrence. The revealed information allows considering the realized cost blending the boundary between collision and non-collision, 
			$l_{nk} = l_{nk}^c   \mathbbm{1}_{c_k}  +  l_{nk}^a   (1-\mathbbm{1}_{c_k})$ where $l_{nk}^c$ is collision-occurred cost\footnote{Collision may alter mean of cost distribution incurred by adversary, e.g., increased-mean distribution, adding extra uncertainty to the decision-maker.} and $l_{nk}^a$ is adversary-selected cost of an agent $n$  pulling an arm $k$. However, such non-arbitrary and synchronous indication information 
			is not valid in an unknown game\footnote{Entities may be reluctant to reveal such indication, and even if it is available, it does not ensure robustness against any potential jamming attackers.}, which makes a multi-agent adversarial MAB setting more challenging due to indistinguishable sources of cost uncertainty.

			This work focuses on an unknown game where such additional indication information is not valid. In practice, an agent can only observe the final (realized) costs. It has no knowledge about the source of the realized cost indistinguishably\footnote{The difficulty lies in that the 0 rewards or 1 cost can indistinguishably come from collisions or null arm capacities.} generated by the oblivious adversary and affected by the congestion degree of the arm $k$ in question. The realized cost sequence can be expressed in a form of the additive outlier model proposed in \cite{Fox1972} as follows: $l_{nk} =  l_{nk}^a +  (l_{nk}^c-l_{nk}^a) \cdot {o_{nk}}$, where $(l_{nk}^c-l_{nk}^a)$ is the magnitude of the outlier and ${o_{nk}}$ is the binary outlier indicator such that ${o_{nk}} = 1$ if the received loss $l_{nk}$ is an outlier, and ${o_{nk}} = 0$, otherwise. The outlier event could occur due to collision and/or adversary. The outlier indicator becomes equivalent to the collision indicator, if ${o_{nk}} = {o_{uk}}, \forall n, u\in\mathcal{N}$. In practice, the agents' sensitivity levels to the indication of an arm are different due to their personalized demands. Also, the agents never know with certainty which one of the events is true in an unknown game, admitting to replacing the discrete indicator with a continuous one. For ease of exposition but without loss of generality, we consider a case where the continuous indication takes an arbitrary value between 0 and 1\footnote{Any arbitrary constant value bounded within (0,1) results in the learning performance bounded by two extreme cases, e.g., by a collision-free but the adversary-selected cost $o_{nk}= 0$ (identical to a case of $c_k = 1$ or $|\mathcal{N}| = 1$) and a collision occurred cost $o_{nk}= 1$.}. This approach allows for the actual sequences received by an agent to shadow the non-stochastic property in an adversarial environment while considering the effect of congestion degree.

		\subsubsection{Connection between learning dynamics and equilibria}
	       	In an unknown game, the candidate actions of agents may differ,       $\mathcal{K}_n \neq \mathcal{K}_{u}~\forall u \!\in\!\mathcal{N}\backslash n$.  Each agent may have its own payoff function and there are no unilateral payoff ties. Thus the game is asymmetric \cite{Ackermann2009}. The existence of NE in the game is easy to establish using an induction approach \cite[Theorem 1]{Feng2017} \cite[Theorem 1]{Milchtaich1996} and a Kakutani fixed point approach \cite{Glicksberg1952}.  It is desirable to design a distributed algorithm with an attractive property of admitting convergence to equilibria under an information-restrictive situation. In general, however, NEs are unlikely to be a realistic prediction of game outcomes: i) it is unclear how agents are expected to coordinate on an equilibrium outcome in games with multiple equilibria, ii) it is unrealistic to assume that all agents in a system will necessarily play strategies that form an equilibrium, iii) in games with unique NE, finding the equilibrium may require computation using global information about the game play, that users may not have access to, and iv) even when equilibria are easy to compute, there is no guarantee, nor immediate motivation for agents demonstrating selfish behavior to converge to them. As a solution concept for the unknown game, we adopt a relaxed NE a similar concept to NE (Definition~\ref{def_ne}). 
			
			\begin{definition}\label{logit-equilibrium}
				($\xi$-equilibrium~\cite{McKelvey1995}) A strategy profile $p'$ is called $\xi$-equilibrium of a game, if it has a bound for marginal payoff loss of each agent, i.e., $|l_n(p_n'; p_{-n}') - l_n(p_n; p_{-n}')|\leq \xi$ for every deviation $p_n \in \mathcal{X}_n, \forall n\in \mathcal{N}$. The strategy $p_n$ is mapped to a fixed score vector through a vector-valued function, $p_n = \Lambda_n(\cdot)$, e.g., a logit choice map in this work. At least one $\xi$-equilibrium exists with logit dynamics \cite{McKelvey1995}.
			\end{definition}
			Note that NEs can be viewed as not just a stable steady-state, but the inevitable result of individual adaptation behavior. The agents can play the game by updating their strategy in a way as to optimize individual objectives selfishly and then reach a $\xi$-equilibrium. That is, the relaxed equilibrium accounts for incomplete information and random perturbation, allowing for better versatility in describing the outcomes of natural dynamic processes in a repeated unknown game as compared to NE. One next question is how to design an algorithm to achieve the relaxed solution taking into account the heterogeneous and dynamic nature of environments
			 
		\subsection{{Perturbed learning dynamics}}
		Next, we design a learning algorithm that uses two exploration processes: assessment and selection. The former describes the way with which agents aggregate their past cost information to update their actions’ scores, and the latter details how these scores are used to select a mixed strategy.
  
            \subsubsection{Perturbed exploration in assessment rule}
		After an agent $n$ selects a suitable VFN $k$ for the upcoming task $\tau$ and offloads it to the selected VFN, it receives a real-valued cost $l_{nk}(\tau)$. Individual assessment rule is used to independently convert the realized cost $l_{nk}(\tau)$ into the learning-weighted estimate of the cost $\hat{l}_{nk}(\tau)$ additive to the previous score representing the VFN's estimated capacity. The individual conversion process is associated with the following explicit and implicit learning parameters. The fluctuations of the loss estimates can be controlled at the price of introducing some bias, alleviating the uncertainty of the empirical estimates in the assessment rule.
			
            \subsubsection*{\textbf{Parameter $\eta'$}}
		A learning rate determines the importance of the estimated cost for $\tau$ in terms of contribution to the cumulative score. It is a parameter controlling how much the weights of the current estimated cost are considered. When the learning rate is large, $p_{nk}$ becomes more uniform, and the algorithm explores more frequently. For a lower learning rate, $p_{nk}$ concentrates on the action with the lowest estimated cost, and the resulting algorithm exploits aggressively. As learning iteration goes on, one agent may want to exploit observations obtained so far to identify the best strategy without engaging others too often. This work considers tuning the learning rate iteration-dependently, exploring less over round; decreasing the learning factor with round. Furthermore, if the exploration-exploitation levels change too fast, it would be too short to obtain the inflection point from exploration to exploitation. For this matter, one may further consider varying the learning factor with the number of candidate sets; the larger the number of actions is, the more slowly the learning factor decreases. Each client may have its own desire for learning rate that are not necessarily the same as other clients. Such unequal learning rates $\eta_n' ({\tau}) \neq \eta_u'({\tau}), {u\in\mathcal{N}\backslash n}$,  could be provoked by two aspects, i) asymmetric arms,  i.e., different clients may be exposed to the different environments, $\mathcal{K}_n(\tau) \!\neq\! \mathcal{K}_{u}(\tau)$, and ii) asynchronous updates, i.e., their learning updates could be performed at individual update time instants and thus the number of active agents could be different over time, $\tau = \tau_n$ and  $\tau = \tau_u$ for clients $n$ and $u$, respectively, where $ \tau_n =\vartheta_n(\tau) =\sum_{\tau'=1}^{\tau}\!\mathbbm{1}_{n\in\mathcal{N}(\tau')}\leq \tau$, thus possibly $\tau_n \neq \tau_u, \forall n,u\in\mathcal{N}$. Such asymmetric and asynchronous nature of the cost estimation update in the assessment rule provokes unequal desires for learning rates.

	\subsubsection*{\textbf{Parameter $\gamma'$}}
	  In a bandit setting, the loss from an action (arm)  $k \neq k'$ could not be   observed due to incomplete feedback. This motivates us to use unbiased estimate, $\hat{l}_{nk}$, that the agent $n$ observe, enabled by i) using the loss $l_{nk}$ if one observes it and $0$ otherwise, $\hat{l}_{nk} = l_{nk} \mathbbm{1}_{{k} = k'}$, and ii) correcting the bias from dividing it by the probability of selecting the action, $\hat{l}_{nk} = \frac{l_{nk}\mathbbm{1}_{k = k'}}{p_{nk}} $, thereby maintaining the expectation property and making actions (arms) that have not been pulled yet optimistic and being likely to be explored. However, such unbiased estimate causes large fluctuation in the loss due to inverse-proportion to $p_{nk}$. One way to change the cost estimates is to control the variance at the price of extra bias. To achieve this, we consider the Exp3 algorithm endowed with implicit exploration (IX)-style cost estimates \cite{Neu2015}. {After each action, the cost is first calculated as $\hat{l}_{nk} = {{l}_{nk}}/({p_{nk} +  \gamma_{n}'})   \mathbbm{1}_{k = k^{'}}$. It is a biased estimator due to $\mathbb{E}[\hat{l}_{nk}] = \sum_{k} p_{nk} \hat{l}_{nk}  = {l}_{nk} -  \frac{\gamma_{n}' \cdot l_{nk'}}{p_{nk'} +  \gamma_{n}'} \leq l_{nk}$ where $ \gamma_{n}'$ is an implicit exploration rate of agent $n$.} The implicit exploration makes the resulting probability of agent $n$ smoother than when using an explicit exploration approach \cite{Neu2015}, i.e., mixing $p_{nk}$ with the uniform distribution. The smoothness property allows actions with large losses to be chosen occasionally, while the actions may render negligible probabilities in the classical algorithm, Exp3. The fluctuations of the loss estimates are not large and thus the estimator is allowed to guarantee reliable performance in rapidly changing, adversarial environments. 		
	\subsubsection{Perturbed exploration in selection rule}
        We incorporate the observation on the resource provider (VFN)'s volatility and resource requester (agent)'s task size into the selection rule in an    adversarial setting, which could select a suitable fog node, rather than a capable one. This achieves a better balance between exploration and exploitation \cite{Cho2021} where the dynamic resource supply and demand-based exploration bonus is augmented in the score $\mathcal{\hat{L}}_{nk}(\tau-1)$ towards a fair and suitable VFN selection.		
	
        \subsubsection*{\textbf{Parameter $\delta$}}
	   Note that while the objective in equation (\ref{eq:P0}) is to optimize the expected unit cost of offloading the given task  $\tau$ to VFN $k$, what actually needs to be learned is the potential capability of each candidate VFN $\mathcal{\hat{L}}_{nk}({\tau-1})$ and its projected suitability for $\tau$ under an adversarial framework. The latter can be evaluated with the normalized end-to-end delay of offloading at a task level, $q_n(\tau)\mathcal{\hat{L}}_{nk}({\tau-1})$. Such joint consideration of both the normalized delay per bit and per task $(1+q_n(\tau)) \mathcal{\hat{L}}_{nk}({\tau-1})$ may take some coordination in terms of input data size-dependent exploration-exploitation trade-off. For the feature scaling, the normalized size of $\tau$ is used as a weight factor ${\zeta}_n({\tau}) = 1+ \delta_n({\tau})$, where $\delta_n({\tau}) = (q_{n}({\tau})-\underline{q}_n)/({\bar{q}_n-\underline{q}_n})$, on the offloading delay in decision-making algorithm \cite{Cho2021}, i.e., $\mathcal{W}_{nk}({\tau}) = e^{- {{\zeta}_n({\tau})}\cdot\mathcal{\hat{L}}_{nk}({\tau-1})}$. Without loss of generality, individually different but temporally fixed demands can be considered, e.g., ${\zeta}_n({\tau}) = {\zeta}_n, \forall n, \tau$.  
			
        \subsubsection*{\textbf{Parameter $\beta$}}
	   If an action (VFN) ${k}\in\bar{\mathcal{K}}_n(\tau')$ appears in round $\tau'$, $\mathcal{K}_n({\tau'}) = \mathcal{K}_n({\tau'-1})\cup \bar{\mathcal{K}}_n(\tau')$ where $\bar{\mathcal{K}}_n({\tau'})$ is the set of the VFNs appearing at $\tau'$, as the previous candidate set of VFNs did, i) all actions including the new action could be reset, $\mathcal{\hat{L}}_{{nk}}({\tau'}) = 0, k\in\mathcal{K}_n(\tau')$ (full reset), or ii) only the new action's score is initialized with zero, $\mathcal{\hat{L}}_{n{k}}({\tau'}) = 0, k\in\bar{\mathcal{K}}_n(\tau')$ (partial reset). However, such conventional resetting mechanisms may invalidate the score based learning benefit in the rapidly changing environment, due to the inefficiency, i.e., the old actions may sacrifice their opportunities, regardless of their accumulated experience, to learn the dynamic task offloading environment. Such an unfair selection rule from the perspective of old actions could be amended by setting the initial score of each appearing action with the already existing one from oneself or others \cite{Cho2021}, e.g., for $k\in \bar{\mathcal{K}}_n({\tau'})$,  $\mathcal{\hat{L}}_{n{k}}({\tau'}) = \mathcal{\hat{L}}_{n{k}}({\tau'}) + \beta_{nk}$, where $\beta_{n{k}} = \max[\mathcal{\hat{L}}_{n{k}}({\tau'-1}), \min_{m\in {\mathcal{K}}_n({\tau'-1})}\mathcal{\hat{L}}_{nm}({\tau'-1}) ]$. 		 
	   Taking into account the above-mentioned perturbations, this learning process can be described in pseudo-code form as in Algorithm \ref{algorithm1}. The playing of all agents is uncoupled, which means that each agent has individual learning parameters annealing with respect to $\tau$. Each agent adopts a regret-based procedure to update its mixed strategy, which depends only on its past costs. According to our previous work \cite{Cho2021}, the corresponding regret has sub-linearity and could be reduced compared to the case without taking into account dynamic resource demand and supply. While regret captures the learning objective of an individual agent, at the system level, it is desirable to know i) whether the dynamical behaviors of distributed agents reach an equilibrium in some sense, and ii) whether the self-interested regret minimization promises a certain level of optimality in terms of social welfare. For the case $\mathcal{K}_n(\tau) \cap \mathcal{K}_u(\tau) = \emptyset~\forall n, u \in\mathcal{N}$, there is no difference between individual and system-level performance. As long as $\mathcal{K}_n(\tau) \cap \mathcal{K}_u(\tau) \neq \emptyset $, the impact of individual behavior on the system-level performance becomes non-trivial. 
			
			\begin{algorithm}[t]\small 
				\caption{MIX-AALTO \cite{Cho2021} in game $\Gamma$}  \label{algorithm1}
				\begin{algorithmic}[1]  
					\State Input: Learning step-size sequences, $ \eta_n'$, $ \gamma_n'$, $\vartheta_n = 0, \forall n$
					\For{$\tau\in \mathcal{T}$}   	
					\For{$n\in \mathcal{N}$}   
					
					\If{$n$ is inactive}  	\Comment{Asynchronous learning}
					\State Continue; 
					\EndIf
					
					\State Set $\vartheta_n \leftarrow  \vartheta_n + 1$			
					\State Set $\eta \leftarrow  \eta_{n}'(\vartheta_n)$, $\gamma \leftarrow  \gamma_n'(\vartheta_n)$ 
					\State Set ${{L}}_{nk} \leftarrow \mathcal{\hat{L}}_{nk}, \beta_{nk} \leftarrow 0, k\in \mathcal{K}_n$  
					\State Set ${\mathcal{K}_n} \leftarrow \mathcal{K}_n(\tau), {\bar{\mathcal{K}}_n} \leftarrow \bar{\mathcal{K}}_n(\tau)$ \Comment{Supply} 	
					\State	Set $\beta_{nk} \leftarrow \max[ {{L}}_{{nk}},\min_{m\in {\mathcal{K}}_n\backslash \bar{\mathcal{K}}_n} {{L}}_{nm}], k\in \bar{\mathcal{K}}_n$
					\State Set $\zeta_n \leftarrow q_n$  \Comment{Demand} 
					\State Set $\mathcal{W}_{nk} \leftarrow  \zeta_n \cdot (L_{nk}+ \beta_{nk}), k\in {\mathcal{K}}_n$ 
					\State Set $p_n \leftarrow \left[\frac{\exp(-\mathcal{W}_{nk})}{\sum_{m} \exp(-  \mathcal{W}_{nm})}\right]_{k\in \mathcal{K}_n }$ 	\Comment{Selection}
					\State Select action $k' \sim p_n$
					\State Receive cost ${l}_{nk'}$ \Comment{Assessment}
					\State Compute $\hat{l}_{nk}  \leftarrow \left[ \frac{{l}_{nk} \cdot \mathbbm{1}_{k = k'}}{p_{nk} +   \gamma}\right]_{k\in \mathcal{K}_n }$
					\State Update scores: $\mathcal{\hat{L}}_{nk} \leftarrow \hat{\mathcal{L}}_{nk}  + \eta  \hat{l}_{nk}, k\in  {\mathcal{K}_n}$
					\EndFor 
					\EndFor 
				\end{algorithmic}
			\end{algorithm}

\section{{Convergence analysis}}
        This section studies system-level performance in terms of the convergence to actual equilibria and PoTA, and shows that our proposed solution on the self-interested behaviors inclining toward robustness and adaptation yields lowering the upper bounds of PoTA, thereby achieving a better balance between individual-level acceptability and system-level efficiency.
			
	\subsection{Convergence to an approximated equilibrium}
	In the following, we study the (asymmetric) replicator dynamics \cite{Taylor1978, Cohen2017} for the game, by analyzing a differential equation expressing a continuum limit of the perturbed update process. For a certain agent, $n\in \mathcal{N}$, the learning procedure in the algorithm $\ref{algorithm1}$ can be represented as the following recursion:
	\begin{equation} \label{recur}
		\left \{\begin{array}{l}
			p_{n}(\tau) = \Lambda_{n}\left(\mathcal{W}_n(\tau-1)\right) = \left[\frac{e^{-\mathcal{W}_{nk}(\tau-1)}}{\sum_m e^{- \mathcal{W}_{nm}(\tau-1)}}\right]_{k\in \mathcal{K}_n},\\
			\mathcal{\hat{L}}_n(\tau)  = \mathcal{\hat{L}}_n(\tau-1) +   \eta_n' (\tau)  \hat{l}_{n}(\tau),
		\end{array}
		\right.
	\end{equation}
	where $\hat{l}_{n}(\tau)  = \left[\frac{{l}_{nk}(\tau)\cdot \mathbbm{1}_{k=k'}}{p_{nk}(\tau) +  \gamma_{n}'(\tau)}\right]_{k\in \mathcal{K}}$, $\eta_{n}'(\tau)$ and $\gamma_{n}'(\tau)$ are calculated as a reference rate $\kappa(\tau)$ \footnote{This is not revealed but commonly taken to be the maximum learning rate among agents for the convergence analysis later.} multiplied with an explicit and implicit exploration parameters, $\eta_{n}$ and $\gamma_{n}$, of agent $n$, expressed as $\eta_{n}'(\tau) = \kappa(\tau)\eta_{n}(\tau)$ and $\gamma_{n}'(\tau) = \kappa(\tau)\gamma_{n}(\tau)$. 
	 
	To ensure the convergence of the strategies induced by the proposed algorithm, we show that equation ($\ref{recur}$) is an asymptotic trajectory for the underlying mean dynamic, i.e., its continuous-time version. To compare the actual sequence of play (discrete) to the replicator dynamics of evolutionary game theory (continuous), we employ the powerful ordinary differential equation (ODE) method in \cite{Benaim1999}. The ODE method leverages the convergence of a continuous-time dynamical system to obtain convergence of the algorithm. We also show how to seek convergence to NEs, which is non-trivial in presence of independently perturbed explorations. Next, the dynamic of an individual's probability distribution over the available actions is given by a perturbed replicator. 
	
	\begin{lemma}\label{replicator} (ODE) Considering the algorithm's update rule, the expected update in the probabilities, by deriving the limit as a parameter, $\kappa(\tau) \rightarrow 0$, is  the following first-order differential equation known as the replicator dynamic: 
		\begin{eqnarray}\label{simple_learning}
			\mathbb{E}_n[\dot{{p}}_{nk}] =    \zeta_n \eta_n  {p}_{nk} \left[ \sum_{m  \in \mathcal{K}}    {l}_{nm}{p}_{nm}  -  {l}_{nk}\right]. 
		\end{eqnarray}
		\begin{proof} 	 
			The proof is in Appendix\ref{ref:replicator}.	
		\end{proof}
	\end{lemma} 
	 
	The dynamics can be intuitively understood as an update mechanism where the probability for a client to choose actions whose expected costs are below average will increase in time, while non-beneficial actions may be gradually abandoned. Given the dynamics, a natural question is whether the linear interpolation of the sequences $p(\tau)$ converges to a fixed point of the replicator dynamics.
	
	We focus on the sequences $p_n(\tau)$ and its linear interpolation which would track the continuous-time version of the learning procedure up to imperfection error\footnote{We stress that our aim is not to rectify the particular model imperfections, but to develop an approach which is of value in the ubiquitous case where model imperfection is not known.}, i.e., the iterates $p_n(\tau)$ are interpolated into a continuous-time process with interpolation intervals associated with learning steps. The ODE method is used to show that asymptotically the iterates follow the path of the mean of ODE. The works in \cite{Ljung1977, Kushner1977, Kushner1984} used the ODE methods to cover general noise processes by the use of average conditions. The main idea is to show that, asymptotically, the noise effects average out so that the asymptotic behavior is determined effectively by that of a mean ODE, i.e., the asymptotic of the iterate sequence is analogue to the one of the interpolated sequence \cite{Kushner1984}. 
	
	Consider that elements of $\mathcal{R}$ are limits of perturbed solutions to the ODE, $\frac{\partial p_n(t)}{\partial t}$. In \cite{Benaim1999}, it has been shown that for $\kappa(\tau) \eta_n(\tau) \rightarrow 0$, all limit points of $p_n(\tau)$ belong to $\mathcal{R}$ and that each element of $\mathcal{R}$ can potentially be a limit point of $p_n(\tau)$ with a nonzero probability, $\lim_{i\rightarrow\infty} d(p_n(\tau), \mathcal{R}) = 0$, i.e., the interpolated process of the sequences $p_{n}(\tau)$ is an asymptotic pseudo\footnote{Owing to the types of averaging methods adopted, the noise can be pseudo rather than true random process.} trajectory (APT) of the solutions of ODE. However, this APT could not be guaranteed for the perturbed replicator dynamic, since the limit trajectory of the ODE may not be concentrated at a single point, $\lim_{\tau\rightarrow\infty} d(p_n(\tau), \mathcal{R}) \neq 0$. To accommodate the issue, one considers pseudo-orbits rather than trajectories, reflecting imperfection error where the iterate remains in a small neighborhood of the limit point for enough time before possibly leaving.
	
	The stability of the mean ODE guarantees that the continuous-time process converges to an asymptotically stable equilibrium point of the perturbed replicator dynamics. While ${p}(t)$ converges with probability one to a bounded invariant or limit set of the ODE, it is not always guaranteed that the largest invariant or limit sets contain points to which convergence clearly could occur. Typically, some properties of a Lyapunov function are used to show that a discretized system has a nearby attracting set. According to the Lasalle principle \cite{LaSalle1960} and Corollary 6.6 of \cite{Michel1999}, the existence of a strictly decreasing Lyapunov function implies that ${p}(t)$ converges to the connected sets of fixed points of the dynamic. A potential function admitted in a game could serve as a Lyapunov function for the stabilization of dynamic systems associated to the game. The potential game with users adopting bandit feedback-based algorithms allows to have a strictly decreasing Lyapunov function \cite{Cohen2017}. Even if the game has not yet reached an equilibrium, the turn of a deviating client will arrive eventually and its action will decrease the potential function. However, such improvement property is not guaranteed in an asymmetric scenario where every client has not only a distinct VFN resource set but also client-specific cost realization. 
	To circumvent this issue we show that the iterates not only eventually stay in the compact recurrent set but that they converge to the limit set of the ODE in that recurrent set. Next, we first characterize the chain recurrent set and asymptotic bias of the perturbed ODE \cite{Tadic}.
	
	\begin{lemma}\label{apt} If noise conditions, {$\lim_{\tau\rightarrow \infty} \kappa(\tau)\eta_n(\tau) = 0$} and  $\sum_{\tau} \kappa(\tau)\eta_n(\tau) = \infty$, $\forall n\in\mathcal{N}$, are satisfied, the iterative process (\ref{recur}) tracks the continuous-time system up to a bounded error $\epsilon<\infty$ and converges to the internally chain recurrent set of the mean-field system.
		\begin{proof}
			The proof is in Appendix\ref{ref:apt}.
		\end{proof}
	\end{lemma} 	
	
	Next, we show the converged limit set contained in every attractor under the logit rule, asymptotically stable for its ODE. 	
	\begin{lemma}\label{stable}
		If $2\zeta\theta < 1$ where $\zeta = max_n \zeta_n$ and $\theta$ is an upper bound for the impact over a client's cost when a single client changes its move 
		for each client $n \in \mathcal{N}$, every strategy $k_n\in \mathcal{K}_n$, and all pairs $k_{-n}$, $k_{-n}' \in \mathcal{K}_{-n}$, then $l(\Lambda(\mathcal{W}))$ is a contraction and its fixed point is asymptotically stable for (\ref{simple_learning}). 	
		\begin{proof}	
			The proof is in Appendix\ref{ref:stable}.
		\end{proof}
	\end{lemma}
	
	\begin{remark}
		(Weakly stable)
		One agent's offloading decision on a VFN causes others to reduce their probability of selecting that VFN henceforth. Such asymmetric behavior is in self-reinforcement driven by the decentralized nature of dynamics in the game, and this can be justified by the symmetry-breaking \cite{Kleinberg2009} implied by spectral properties of the Jacobian of the dynamics defined at an equilibrium of the game, i.e., a weakly stable equilibrium in which the Jacobian is allowed to have eigenvalues whose real part is at most a small positive real, i.e., less than a unity \cite{Cho2016}. Thus, agents are able to steer clear of undesirable equilibria in the presence of arbitrary events captured in a discretized time frame\footnote{The time scale is determined by the iteration $\tau$ and the step size $\kappa\eta$. Such interpolated time intervals are natural choices for the problem related to the characterization of the asymptotic properties of the sequences \cite{Kushner2003}.}, i.e., unknown arrivals with different delay values.
	\end{remark}
	
	\begin{corollary}\label{corollary1}
		In a game with a $\theta$-Lipschitz linear cost function w.r.t $c_k$, if $\theta\zeta/2 < 1$, the corresponding dynamics converge to an asymptotically stable fixed point. 		
		\begin{proof}
			The proof is in Appendix\ref{ref:corollary1}.			  
		\end{proof}	
	\end{corollary}
	
	\begin{remark}
		(Asymptotically stable) While $\zeta <2$, the condition $\theta < 1$ is sufficiently satisfied if each VFN $k$ can allocate to all agents selecting the VFN $k$ at least the amount of CPU frequency $f_{nk}$ (cycles/second) larger than the computational complexity $w_n$\footnote{It may vary with applications} (cycles/bit) required for processing one input data bit, e.g, $\theta = \frac{w_n}{f_{nk}} =  \frac{w_n}{w_n + f_{nk}'} < 1$ where ${f_{nk}'}$ is a positive computing capability value of a VFN $k$ further allocated to an agent $n$ than the minimum, $f_{nk} = w_n + f_{nk}'$.  
	\end{remark}
		
	With the fulfilled conditions above, the iterative process (\ref{recur}) tracks the continuous-time system up to a bounded error $\epsilon<\infty$ and, from any initial state, converges almost surely to the stable fixed point of the dynamics (\ref{simple_learning}). One natural question is whether the converging point is consistent with a NE of the game $\Gamma$, stated in Definition \ref{logit-equilibrium}.
	
	\begin{proposition}\label{prop1_convergeNE} (Converged NE)
		If $\sum_{\tau} \kappa(\tau)\eta_n(\tau) \rightarrow \infty$ and $\kappa(\tau)\gamma_n(\tau) > {\tau}^{-1}$, the actual profile sequence $p(\tau)$ converges (a.s) toward a NE of the game $\Gamma$ and the converging point is $\xi$-equilibrium with $\xi = \max_{n\in\mathcal{N}}(\log(|\mathcal{K}_n|)/\zeta_n)$.
		
		\begin{proof} 
			The proof is in Appendix\ref{ref:prop1_convergeNE}.			
		\end{proof}
	\end{proposition}
	
	\begin{proposition} \label{prop:convergencerate} (Converging rate)	The convergence occurs at a quasi-exponential rate,  
		\begin{eqnarray}		
			\Lambda_{n k'}(T) {\geq} {1} - {{(K-1) e^{-\zeta_n\cdot[\Delta_{\beta} +   \Delta_{l} \cdot  \sum_{{\tau}=1}^T  \kappa(\tau)\eta_n(\tau) ]}}},\nonumber
		\end{eqnarray}
		where $\Delta_{l} \leq {l}_{nk}(\tau)- {l}_{nk'}(\tau) \forall \tau$ and $\Delta_{\beta} = \beta_{nk}(0)-\beta_{nk'}(0)$.
		\begin{proof}
			The proof is in Appendix\ref{ref:prop:convergencerate}.
		\end{proof}
	\end{proposition}

	\begin{remark} 
		The game converges faster when i) the number of candidate VFNs, $K$, gets fewer, ii) the minimum cost difference between costs of a selected VFN, $k'$, and other VFNs, ${l}_{nk}(\tau)- {l}_{nk'}(\tau)$, gets larger, and iii) the time period a selected VFN $k'$ has resided in the candidate VFN set of a client, $n$, is longer than or equal to the other VFNs.
	\end{remark}

	\begin{remark} 
	   The game converges faster for the larger resource demand and learning rate, i.e., when $\delta_n$ and $\sum_{{\tau}=1}^T\kappa(\tau)\eta_n(\tau), \forall n$, get larger.
	\end{remark}

	To address the unknown game setup, this work considers strategy learning with an asynchronous update. Due to agents' inherent dynamicity and heterogeneity, the number of active agents may not be constant over time, and no agent knows when the other agents will be active. In the following, we try to make the convergence analysis on approaches used for strategy learning applicable in the asynchronous setting where the set of active agents is variable and unknown\footnote{When the time is divided into several intervals, each client can be run independently each of these intervals as suggested in \cite{Rosenski2016}.}. For the sake of the aim, one may consider an individual clock, a random variable representing the number of times the agent $n\in \mathcal{N}$ has been involved in an interaction until $\tau$, expressed as  $\vartheta_n(\tau) = \sum_{\tau'=1}^{\tau}\mathbbm{1}_{n\in\mathcal{N}(\tau')}$ interacting agents. Rather than requiring a global timer, it is adequate for the analysis above only to consider a reference type of learning rate which refers to the maximum value among the learning rates of the active agents, represented by $\kappa^*(\tau) = \max_{n\in \mathcal{N}(\tau)} \kappa(\vartheta_n(\tau))\eta_n(\vartheta_n(\tau))$.
			
	\begin{proposition}\label{async} (Async update)
	If $\sum_{\tau} \kappa(\tau)\eta_n(\tau) = \infty$ and $\sum_{\tau} \kappa^2(\tau)\eta_n^2(\tau)<\infty, \forall n$, we have $\sum_{\tau} \kappa^*(\tau) = \infty$ and $\sum_{{\tau}} [\kappa^*(\tau)]^2<\infty$.
	\begin{proof}
		The proof is in Appendix\ref{ref:async}.
	\end{proof}
	\end{proposition}

	\begin{remark}
		Asynchronous updates $\vartheta_n(\tau) \neq \vartheta_u(\tau), \forall n,u\in\mathcal{N}$ exhibit inherently aligned and independently calibrated with a referral learning rate $\kappa^*(\tau)$ unknown. Since $\kappa^*(\tau)$ is the largest one, the individual conditions for the convergence are sufficiently satisfied, {that is, the convergence analysis for the case $\vartheta_n(\tau) = \vartheta_u(\tau) = \tau, \forall n,u\in\mathcal{N}$ is valid.} 
	\end{remark}
	  
	\subsection{{Efficiency} } 
	At the system level, one may want to know if the self-interested objective optimizations of multiple clients in $\mathcal{N}$ in equation (\ref{eq:P0}) collectively promise a certain level of optimality in terms of social welfare. We now address the important question of how far the system performance induced the learning algorithm would be from optimal in a NE. The relation between the delay cost induced by such an equilibrium state converged by the learning dynamic (Algorithm \ref{algorithm1}) and the socially optimal solution minimizing the total cost over different clients has been well studied under the PoA \cite{Roughgarden2015}. 
			
	The PoA indicates the suboptimality caused by selfish behavior. The PoA close to one means that the negative impact of selfish behavior is relatively small, all NEs are near-optimal, and hence any equilibrating learning dynamics suffices to reach approximately optimal system performance. The lower bound on PoA is meaningful only if participants can successfully reach an equilibrium. However, individuals might fail to coordinate on a particular equilibrium or fail to compute a NE, which motivates to adopt robust bounds on PoA. The work in \cite{Roughgarden2015} identifies that such efficiency loss can be bounded whenever clients minimize their regrets and the game is $(\lambda, \mu)$-smooth, if, for a NE ${p}'$ and the optimal strategies ${p}^*$, it satisfies the following relation: $C' = \sum_{n\in\mathcal{N}} l_{nk'} = \sum_{n\in\mathcal{N}} l_n(k'; {k}_{-n}') \leq  \sum_{n\in\mathcal{N}} l_n({k}^*; {k}_{-n}') \leq \lambda   C^* + \mu   C'$ where $k^*$ is the socially optimum action of agent $n$, or equivalently $ {{C'}}/{C^*} \leq \lambda/(1-\mu)$ with $\lambda>0$ and $\mu <1$. The relation holds for every pair of strategies, not just NEs or social welfare minimizing outcomes since the smoothness arguments imply worst-case bounds beyond the supermum of PoA with $\inf\{\lambda/(1-\mu)\}$. 
		 
	Next, we generalize the well-known results of \cite{Roughgarden2015}, for different constants $\lambda$ and $\mu$ based on the information available at a time instant, $t$, showing that such optimal welfare can be approached when clients minimize their regrets.

	\begin{proposition} \label{pota_bound} 
	Assume that $R_n(T)$ be the individual regrets up to $T$ for different clients, $n\in\mathcal{N}$, and $\Gamma$ is $\{\lambda(\tau), \mu({\tau})\}$-smooth at each time ${\tau}$, then the efficiency loss of the evolutionary process, also known as PoTA \cite{Xiao2020}, is $\mbox{PoTA}  \leq \rho  + {\sum_{n\in\mathcal{N}} \frac{R_n(T)}{T (1-\mu')  {C^*}}}$ where $\rho  =  {\lambda'}/{(1-\mu')}$ is the robust \mbox{PoA}, $\lambda' = \max_{\tau\in\mathcal{T}} \lambda(\tau)$, and $\mu' = \max_{\tau\in\mathcal{T}} \mu(\tau)<1$. 
	\begin{proof} 
	 The proof is in Appendix\ref{ref:pota_bound}.
    \end{proof}
	\end{proposition}

	\begin{remark}
	For any strategies in the evolutionary process, the worst-case bound generated through smoothness arguments $\rho = \lambda'/(1-\mu')$ is larger than the supermum of \mbox{PoTA}.
	\end{remark}

	\begin{remark}
	The convergence of the \mbox{PoTA} to the \mbox{PoA} of the stage game implies that no-regret learning can fully null the impact of the unknown nature of the game on social welfare.
	\end{remark}
			
	If the regret is sub-linear w.r.t $T$, for any $T\geq {\tau}_o$, $n\in\mathcal{N}$, there exists a non-increasing per-round regret function such that $\lim_{T\rightarrow\infty} \sum_n R_n(T)/T  \rightarrow \epsilon_T \geq 0$. For any almost sure no-regret sequence, as $T$ goes to infinity, $\mbox{PoTA} \leq \rho$ almost surely. When the action profiles are generated by Algorithm \ref{algorithm1} with $\kappa({\tau})\eta_n({\tau})>{\tau}^{-1}$, $\kappa({\tau})\gamma_n(\tau)>{\tau}^{-1}$ and $\gamma_n(\tau)/\eta_n(\tau)\leq 0.5$, then for any $T\geq {\tau}_o$ such that $\mbox{PoTA}  \leq \rho + \epsilon_T/\left[({1 -\mu'}) \sum_{\tau} C_{\tau}^*/T\right]$ almost surely, where $\epsilon_T \rightarrow 0$ as $T \rightarrow \infty$.
			
	\begin{remark}
	The upper bound of {PoTA} induced by Algorithm \ref{algorithm1} with $\delta_n > 0$ and $\beta_{n} >0$ is lower than the one with $\delta_n = 0$ and $\beta_{n} = 0, \forall n$, where $\beta_{n} = \sum_k \beta_{nk}$. 
	\end{remark}

	\section{{Numerical illustration}}
	This section conducts numerical studies to show the converging sequences of dynamics and their efficiency in terms of PoTA.  For the simulation evaluation, we follow the detailed simulation setting as in \cite{Cho2021}.
			
	\subsection{{Evaluation setting}}
	Consider multiple clients of interest, requesting the computational resource from candidate edge computational resource-providing vehicles (VFC nodes). Three different clients are considered, $|\mathcal{N}| = 3$. The distance between the client and each candidate VFC node is assumed to follow a uniform distribution, $d\sim\mathcal{U}[0,d_r]$ where $d_{r}$ is the communication range equal to 400 m. The transmission power of the client is $24$ dBm, the large-scale fading gain follows the 3GPP pathloss model~\cite{3GPP2011}, $A_o = 128.1 + 37.6\log_{10}(d)$, the small-scale fading gain follows Rayleigh distribution with unit variance, channel bandwidth is $W = 10$ MHz, the number of subchannels is 10, and noise power is $N_o = -174$ dBm/Hz. A subchannel is equally divided by the number of agents and the noise power for an agent is proportional to the bandwidth of interest. 
			
	Consider 10 volatile VFNs $\mathcal{K} = \{1, \cdots, 10\}$ with the respective maximum CPU frequency values, $F_k \in \{6, 6, 5, 4, 1.5, 2, 4, 6, 4, 5\}$ GHz that appear or disappear as candidate fog nodes of one task generating client for a finite number of time frames  (tasks) in 3 epochs, within each epoch consisting of 1000 tasks and keeping the same fog node set for a client, which could be identical or different for different clients. For each VFN $k$, the allocated CPU frequency to the task client $n$, $f_{nk}$, is a fraction of the maximum CPU frequency which is distributed from 20\% to 50\%, but arbitrarily constrained by an adversary, e.g., affected by the oblivious attack as in \cite{Cho2021}, and by the resource congestion due to some or all other clients offloading to the same VFN, e.g., affected by the number $c_k$ of clients selecting the VFN $k$, both of which could yield collision-free but adversary-selected cost sequence, $l_{nk}^a$, and collision occurred cost sequence, $l_{nk}^c$, of the task client $n$, respectively. The available computing resources of a VFN $k$ are non-linear decreasing w.r.t to the total number, $c_k$ of clients offloading to the same VFN $k$ \cite{Gummadi2013}, e.g., $1/\sqrt{c_k}$. The realized task offloading cost is assumed to be from any linear combination of the two sources, $l_{nk}^a$ and $l_{nk}^c$ with ${o_{nk}}\sim \mathcal{U}[0,1]$\footnote{It approximately settles down to the case of $o_{nk} = 0.5$.}, $l_{nk} =  l_{nk}^a (1 - {o_{nk}})+   l_{nk}^c  \cdot {o_{nk}}$.  The total tasks are split into phases with different lengths, each of which is with different means for different arms. The computation intensity is set to $w = 1000$ Cycles/bit. To meet the client's diverse demands, the request service type can be changed with different task sizes arbitrarily. Varying service types could be considered at regular intervals. For simplicity, a periodic interval for changing service types is aligned with an epoch. The task size, $\delta$ Mbits, is either fixed or randomly distributed according to either a uniform or truncated normal distribution on a predefined interval $\delta \in [0.2, 1]$.  
	 
	\begin{figure}[t]
				\centering
				\begin{tabular}{c c} \hspace{-15pt} 
					\subfigure[\tiny \label{numerical:all:benchmark}]
					{\includegraphics[width=0.265\textwidth]{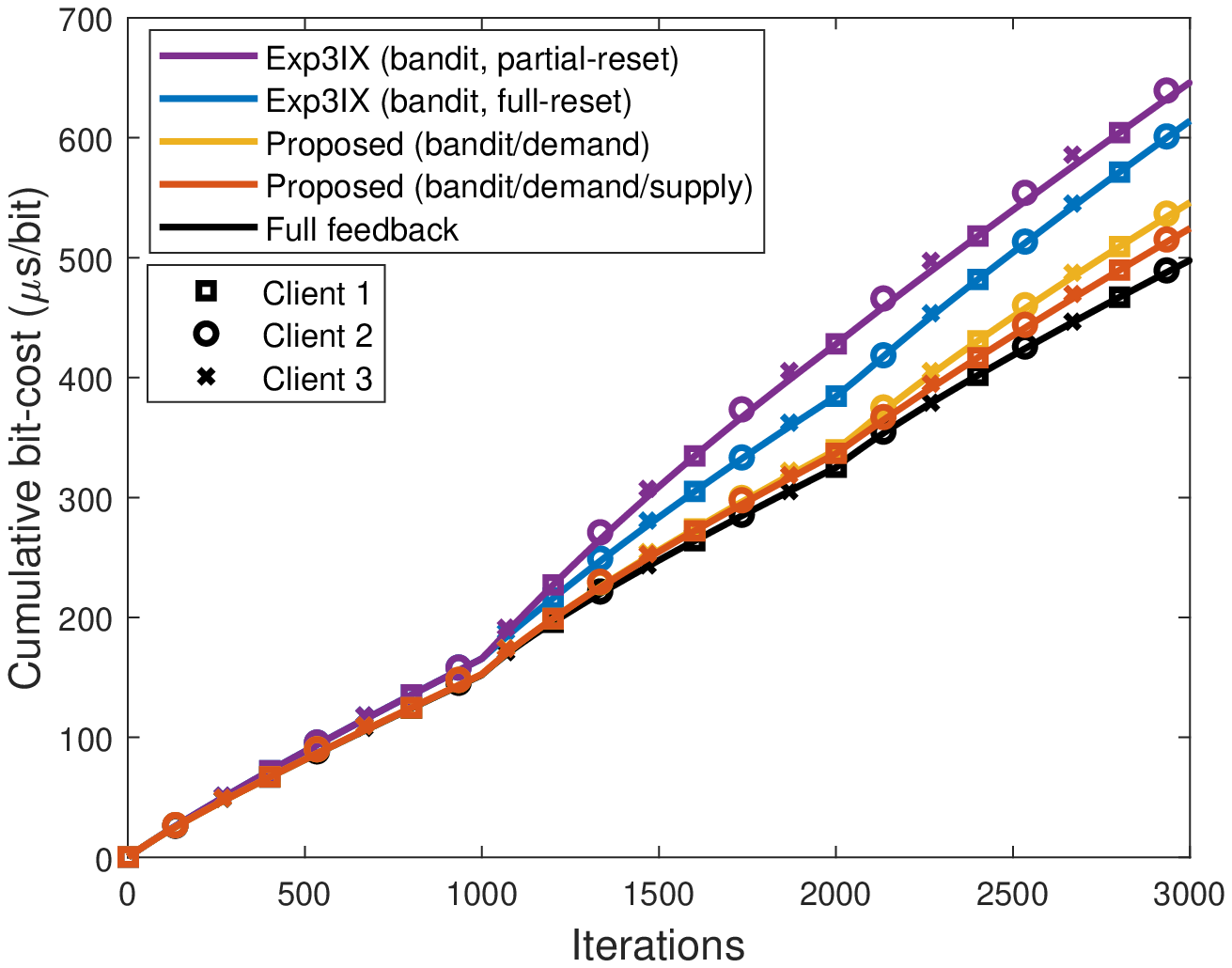}} & \hspace{-2.2em}
					\subfigure[\tiny \label{numerical:all:pota}]
					{\includegraphics[width=0.265\textwidth]{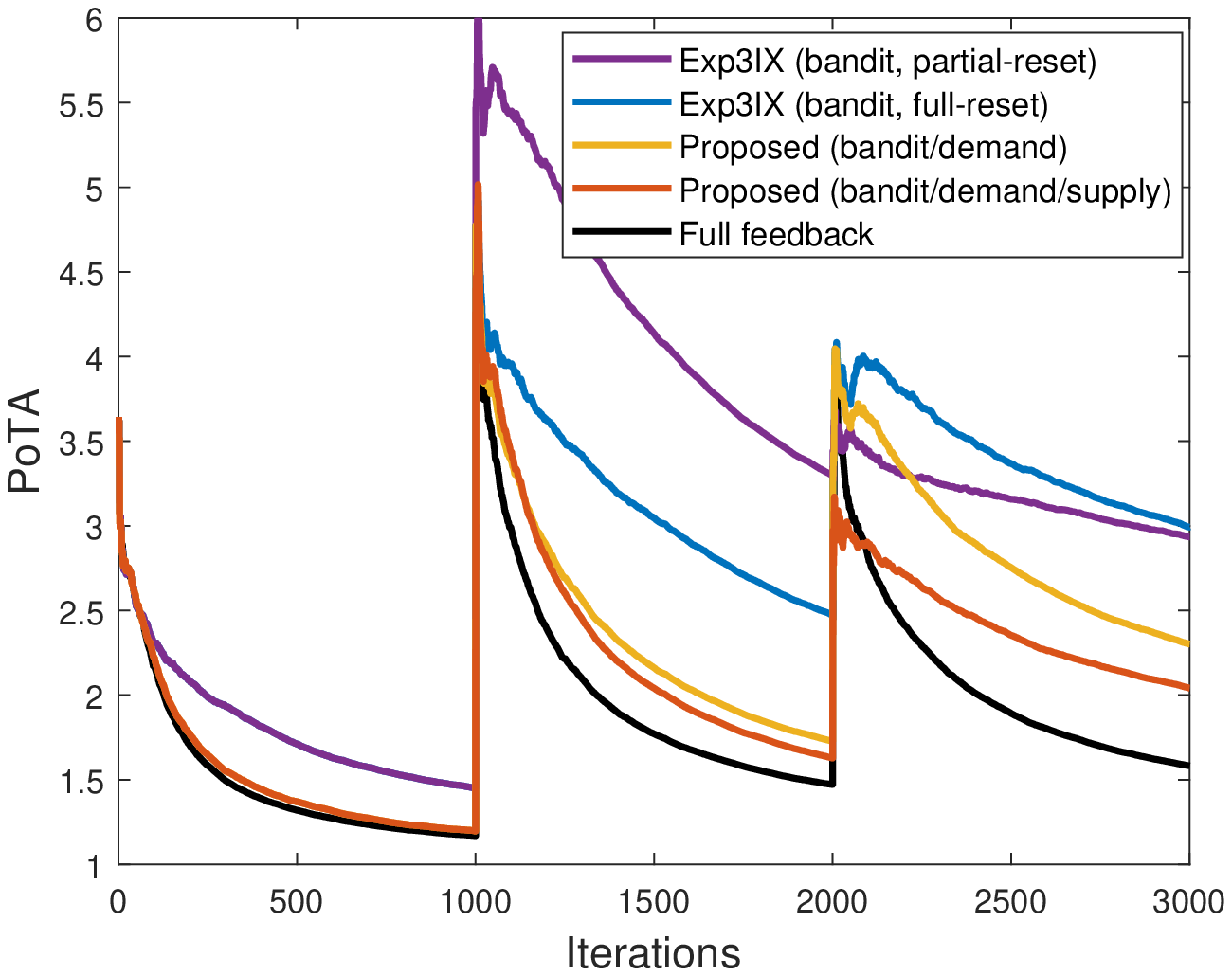}} 
				\end{tabular}\vspace{-.5em} 
				\caption{Impact of resource demand ($\delta>0$) and supply ($\beta>0$) dependent VFN selection on the convergence performance: (a) cumulative bit-cost and (b) PoTA, when $\mathcal{K} = \mathcal{K}_n(\tau), \forall n\in \mathcal{N}$, $\mathcal{K} \!= \!\{1,2,3\}$ for $\tau \!=\! [1,1000]$,
					$\mathcal{K} \!=\! \{1,\cdots,5\}$ for $\tau \!=\! [1001,2000]$, and 
					$\mathcal{K} \!=\! \{1,\cdots,10\}$ for $\tau \!=\! [2001,3000]$.  
				} \label{numerical:all}
			\end{figure}
			
			\begin{figure}[t]
				\centering
				\begin{tabular}{c c} \hspace{-15pt} 
					\subfigure[\tiny \label{numerical:demand:benchmark}]
					{\includegraphics[width=0.265\textwidth]{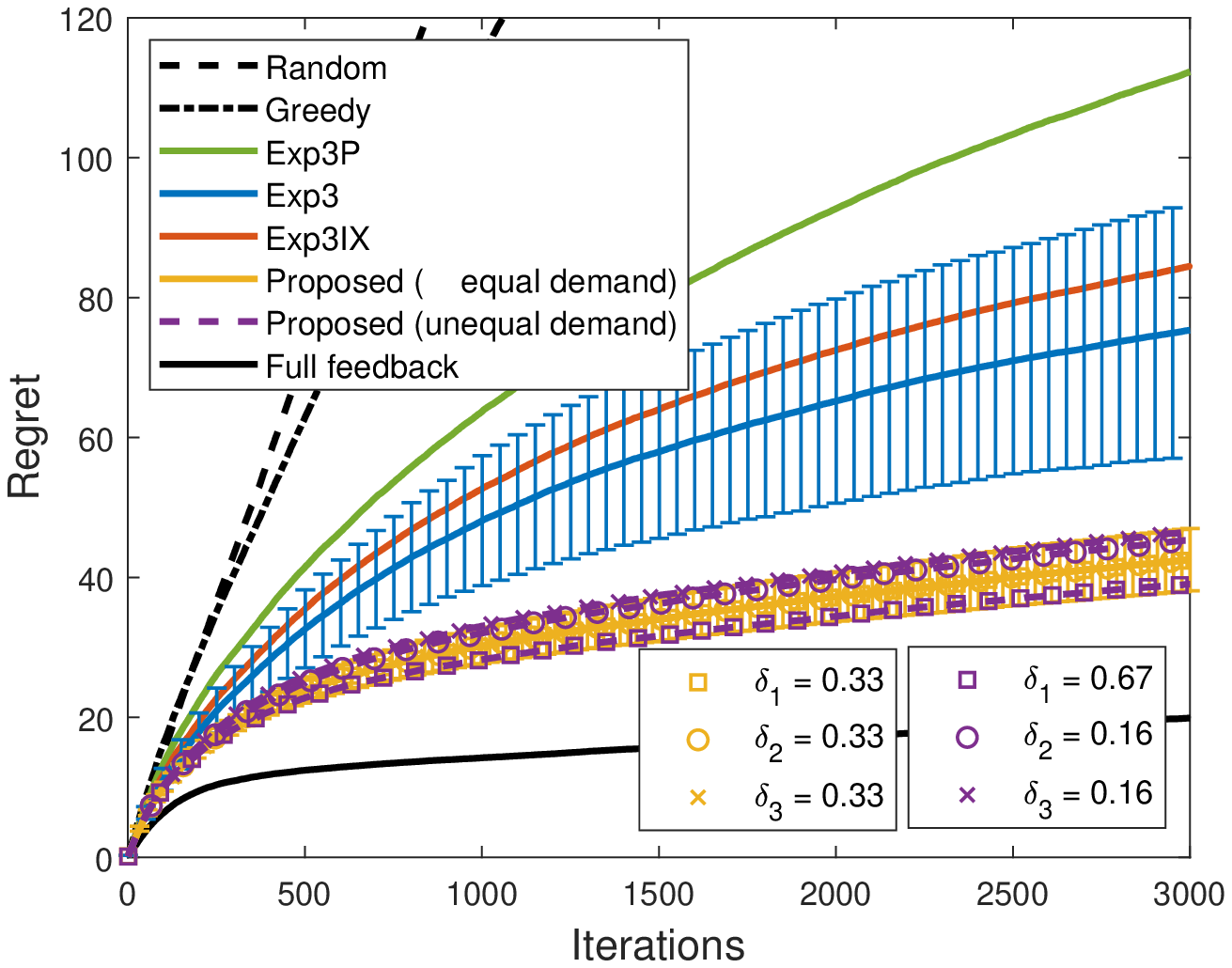}} &
					\hspace{-2.2em}
					\subfigure[\tiny \label{numerical:demand:regret}]
					{\includegraphics[width=0.265\textwidth]{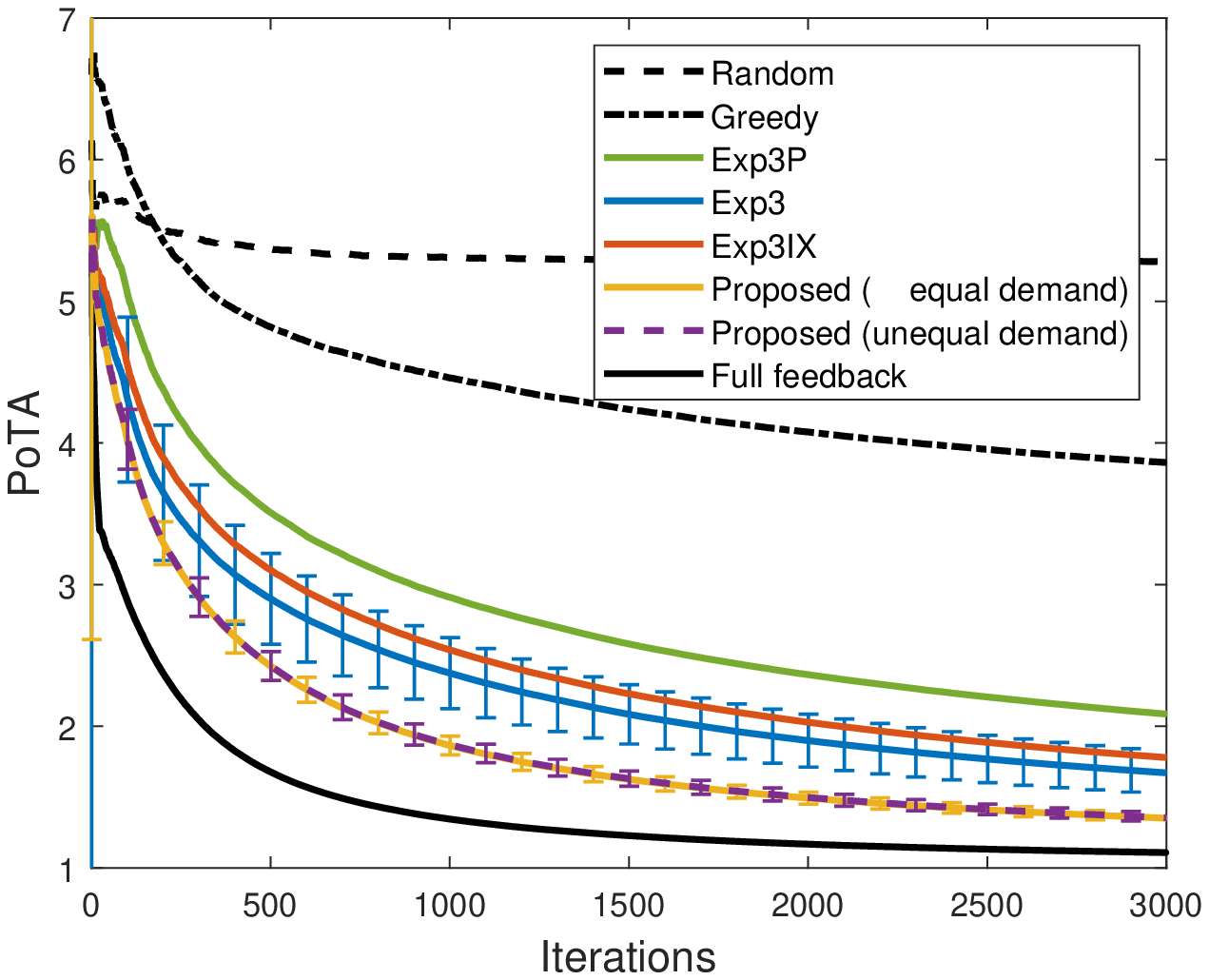}} \\
				\end{tabular}\vspace{-0.5em}
				\caption{Impact of suitability-based VFN selection on the regret and PoTA, when $\mathcal{K} = \mathcal{K}_n(\tau), \forall n\in \mathcal{N}$, $\mathcal{K} = \{1,\cdots,7\}$.} \vspace{-1em}
				\label{numerical:demand}
			\end{figure}

	\subsection{{Evaluation result}}
		
	\subsubsection*{{Benchmark}}
	The perturbed learning algorithm is compared with its counterparts, implicit exploration-based algorithms. The performance results of learning algorithms in terms of the cumulative bit-cost summed over the clients with $|\mathcal{N}| = 3$, and PoTA, are depicted in Fig. \ref{numerical:all}, showing that the proposed algorithm exhibits converging behavior and outperforms other implicit exploration-based algorithms where an arm is selected based on the scores $\hat{\mathcal{L}}_{n}$. Every computing resource client using learning strategies perturbed with $\zeta_n (\hat{\mathcal{L}}_{n} + \beta_{n})$ where $\beta_n>0$ and/or $\delta_n>0$ could achieve a better exploration-exploitation trade-off, since $\beta_n>0$ could save unnecessary exploration time for the arms appearing to the available VFN set of a client $n$ and $\delta_n>0$ could add more input-size dependent importance weight on the arms having the better scores. Such adaptive perturbation strategies allow to reduce the sum bit-cost by 11\% in Fig. \ref{numerical:all:benchmark} and the PoTA by 17\% in Fig. \ref{numerical:all:pota} from that of the vanilla Exp3IX algorithm, and being much closer the full information setting where the complete cost vector is revealed after every round (full feedback). As the number of epochs increases, the sum bit-cost resulting from the perturbed learning algorithm decreases, while the vanilla algorithm increases. This is because the patching rule-based scores enforce the accumulated experience even in the evolving circumstance, while the vanilla one may not be supportive. Note that, however, no such opposite phenomenon occurs in terms of PoTA, since the increasing number of VFNs increases the robust PoA bound $\rho$.
			
	\subsubsection*{{Impact of $\delta$}}
	A task requester has personalized task resource demand affecting $\zeta_n$, $n\in\mathcal{N}$, which can be considered for drawing suitability-based selection in the exploration process of the online learning algorithm. Fig.~\ref{numerical:demand} shows its impact on regret and PoTA. When a positive value of the normalized input data size $\delta_n>0$, $n\in\mathcal{N}$ is considered in a selection rule, a client's learning regret performance can be improved compared to others including vanilla Exp3, Exp3P, and Exp3IX algorithms, since i) considering a score associated with both normalized per-task cost and per-bit cost, make a more suitable candidate and thus ensure a better trade-off between exploitation and exploration, and ii) an implicit exploration approach $\kappa\gamma_n>0$ could achieve better and more robust performance in terms of regret due to lower empirical mean and standard deviation of the regret than others \cite{Cho2021} and thus of PoTA (Prop.~\ref{pota_bound}). The better regret performance gain can be achieved by making exploitation more for a large $\delta$ and less for a small $\delta$. For example, the per-bit learning regret of a client with $\delta_1$ equal to $0.67$ is lower than the ones of others with $\delta_{2}$ and $\delta_{3}$ equal to $0.16$. The resource demand dependent algorithm brings unequal gains in terms of regret for the heterogeneous tasks resource demands, while it results in more or less the same gains in terms of system-level performance, PoTA, if those cumulative demands are homogeneous one another. The excessive surplus earnings from a task of large size and less surplus ones from a task of small size can be offset in social welfare increment, thereby making the upper bounds of PoTA for the different combinations of resource demands but with the same aggregate demand over clients. 
			
   \begin{figure}[t]
				\centering
				\begin{tabular}{c } \hspace{-15pt} 
					\subfigure[\tiny \label{numerical:demand:benchmark}]
					{\includegraphics[width=0.3265\textwidth]{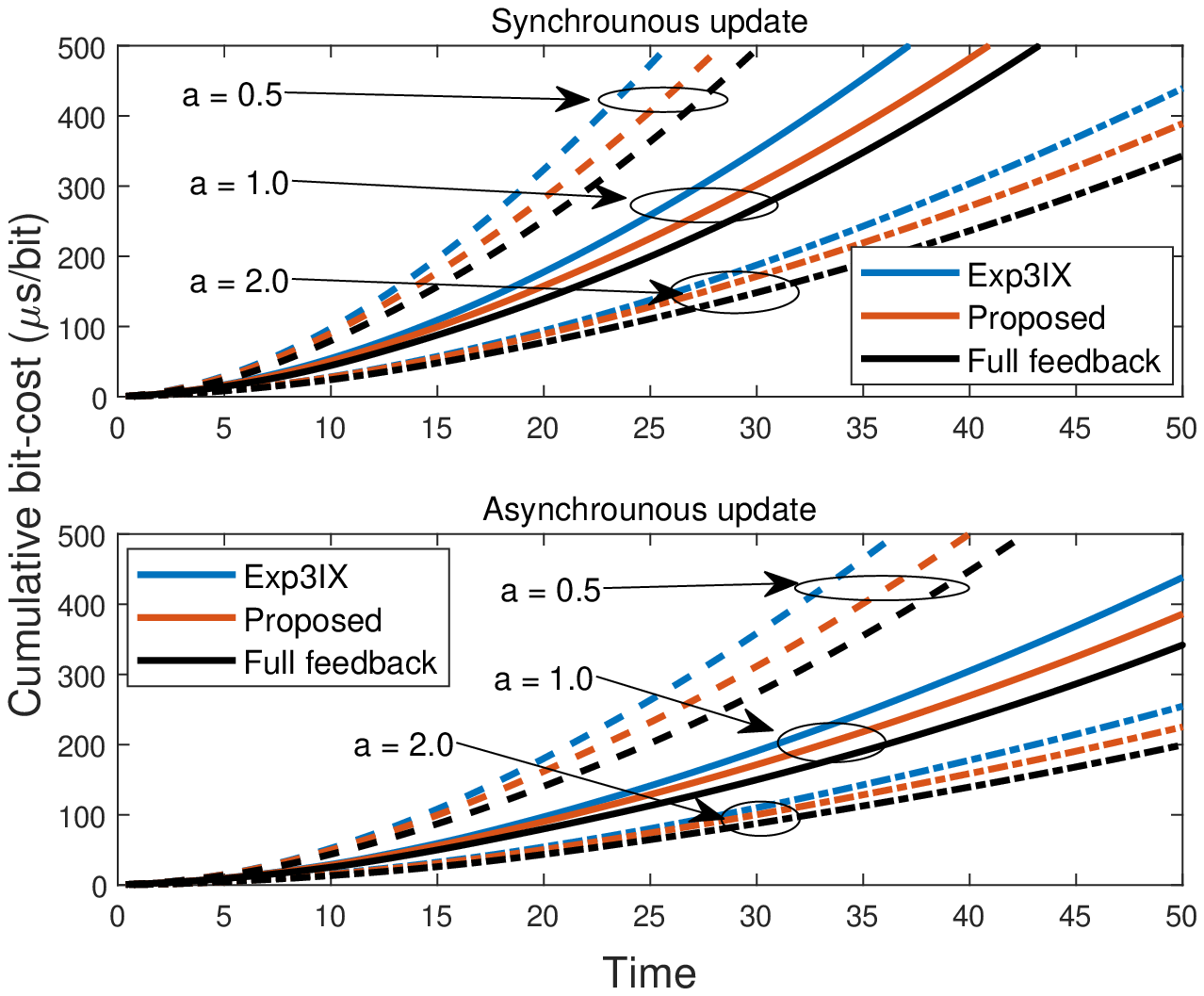}} \\
					\hspace{-20pt}
					\subfigure[\tiny \label{numerical:demand:regret}]
					{\includegraphics[width=0.3265\textwidth]{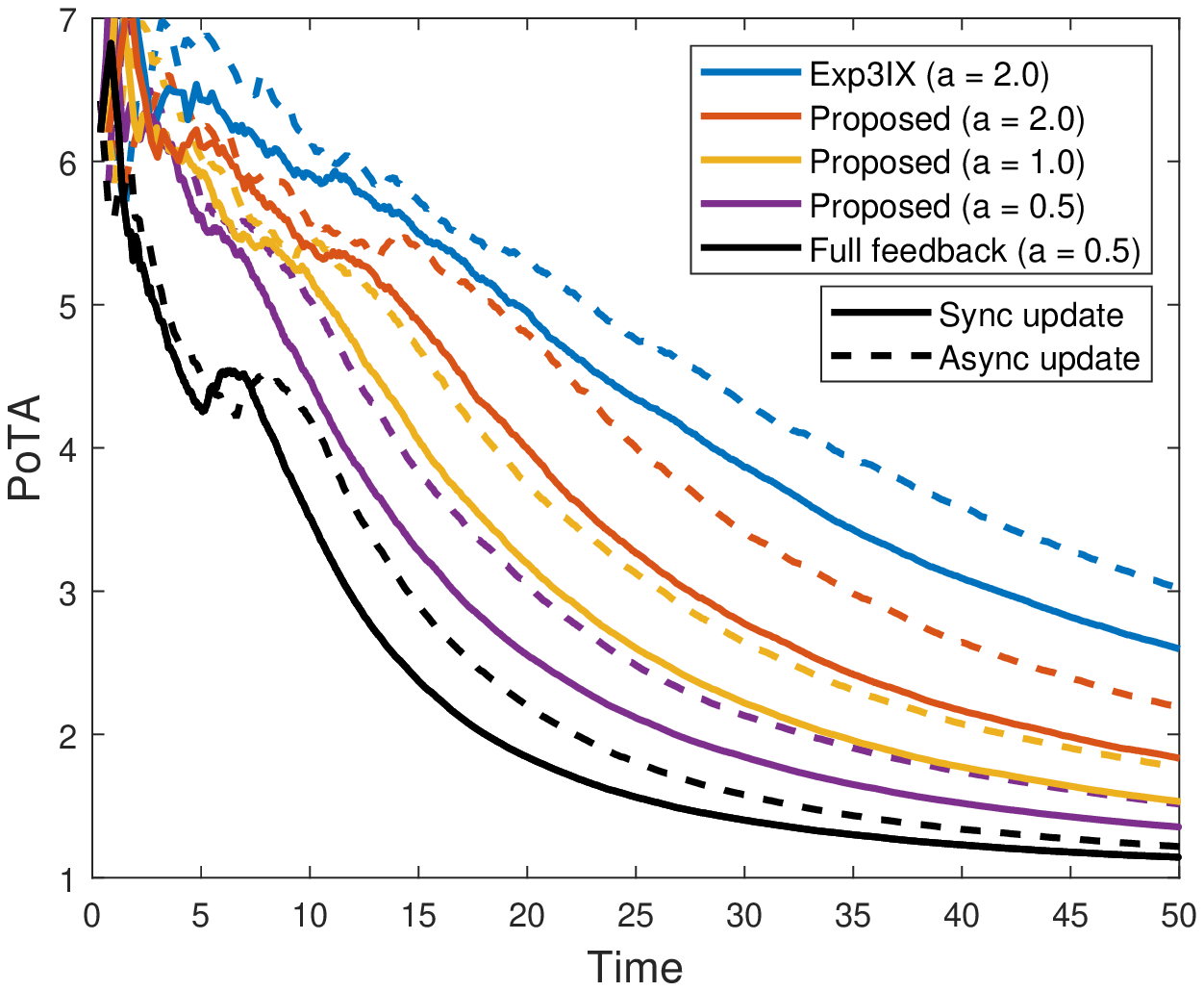}} 
				\end{tabular}\vspace{-.1em}
				\caption{Impact of personalized learning rates $\kappa({\tau})\eta_n({\tau}) = \sqrt{\frac{a\cdot\log(|\mathcal{K}_n|)}{|\mathcal{K}_n|\tau}}$ on the cumulative bit-cost (asymmetric $a$) and PoTA (symmetric $a$) w.r.t $a>0$ referring to different learning speeds: slow, medium and fast learners.} \vspace{-1em}   
				\label{numerical:asymmetric_alpha}
			\end{figure}
			
			\subsubsection*{{Impact of $\kappa\eta$ and $\vartheta$}}
			Fig.~\ref{numerical:asymmetric_alpha} shows the effect of the personalized learning rates on the learning regret and PoTA. A resource client has individual learning rate $\kappa\eta_n$, $n\in\mathcal{N}$ used for piling up desired cost estimation, time-varying importance of the estimated cost, $\kappa({\tau})\eta_n({\tau})\hat{l}_n({\tau})$, to an accumulated score, $\hat{\mathcal{L}}_n({\tau}-1)$. In practice, the learning rates, $\kappa({\tau}) \eta_n({\tau})$, are not necessarily the same for different clients who have a common tendency to assume the decreasing learning rate with round; the more distant the past, the more its learning factor, e.g., $\kappa({\tau})\eta_n({\tau}) = \sqrt{\frac{a\cdot\log(|K_n|)}{|K_n|\tau}}$ \cite[Corollary 2]{Cho2021} where $\kappa({\tau})$ is the unknown referral learning rate among the clients for the assumed attribute over the sampled interval\footnote{Every client does not know the global clock, but only knows how many time she has been active, the activity of others is not known.} of the continuum that is of interest. One may observe that a client with a larger learning rate or more slowly decreasing learning rate in conjunction with iterate, learns at a faster rate (Prop. \ref{prop:convergencerate}), effectively speeds up a learning process, but its individual learning performance in terms of bit-cost is vanishing too fast, resulting in a lower cumulative bit-cost in Fig.~\ref{numerical:asymmetric_alpha}(a). This phenomenon can be also validated in asynchronous updates $\vartheta_n(\tau) \neq \vartheta_m(\tau)$, which lower bounds on the synchronous case due to $T\geq \vartheta_n(T)$ and obtains the similar result in the synchronicity but with a doubled value of $a$ due to independently and uniformly randomized activation. The achieved strategy could be suboptimal due to \cite[Remark 7]{Cho2021} prone to premature convergence, Fig.~\ref{numerical:asymmetric_alpha}(b). The PoTA of clients with a larger learning rate is larger than the one with a smaller learning rate. This phenomenon is due to the fact that the personalized learning rate affects the algorithm’s efforts between exploration and exploitation: a smaller learning rate leads to more conservative exploration over the candidate action set, $k\in \mathcal{K}_n$, while a larger learning rate leads to more aggressive exploration but with insufficient time to learn.	 
			
			\begin{figure*}[t]
				\centering
				\begin{tabular}{c  c c c} \hspace{-15pt} 
					\subfigure[\tiny \label{numerical:demand:benchmark}]
					{\includegraphics[width=0.265\textwidth]{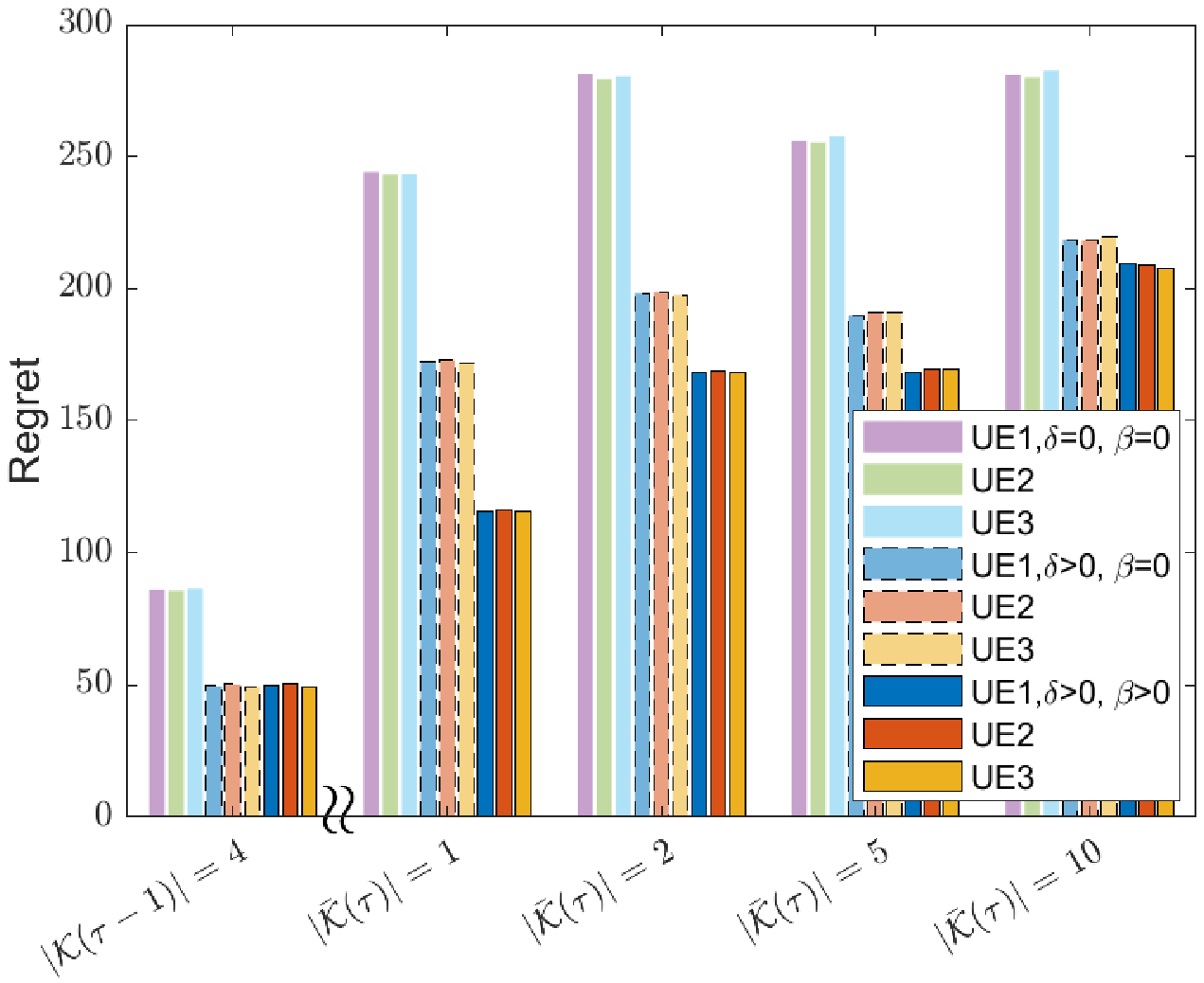}}&
					\hspace{-20pt}
					\subfigure[\tiny \label{numerical:demand:regret}]
					{\includegraphics[width=0.265\textwidth]{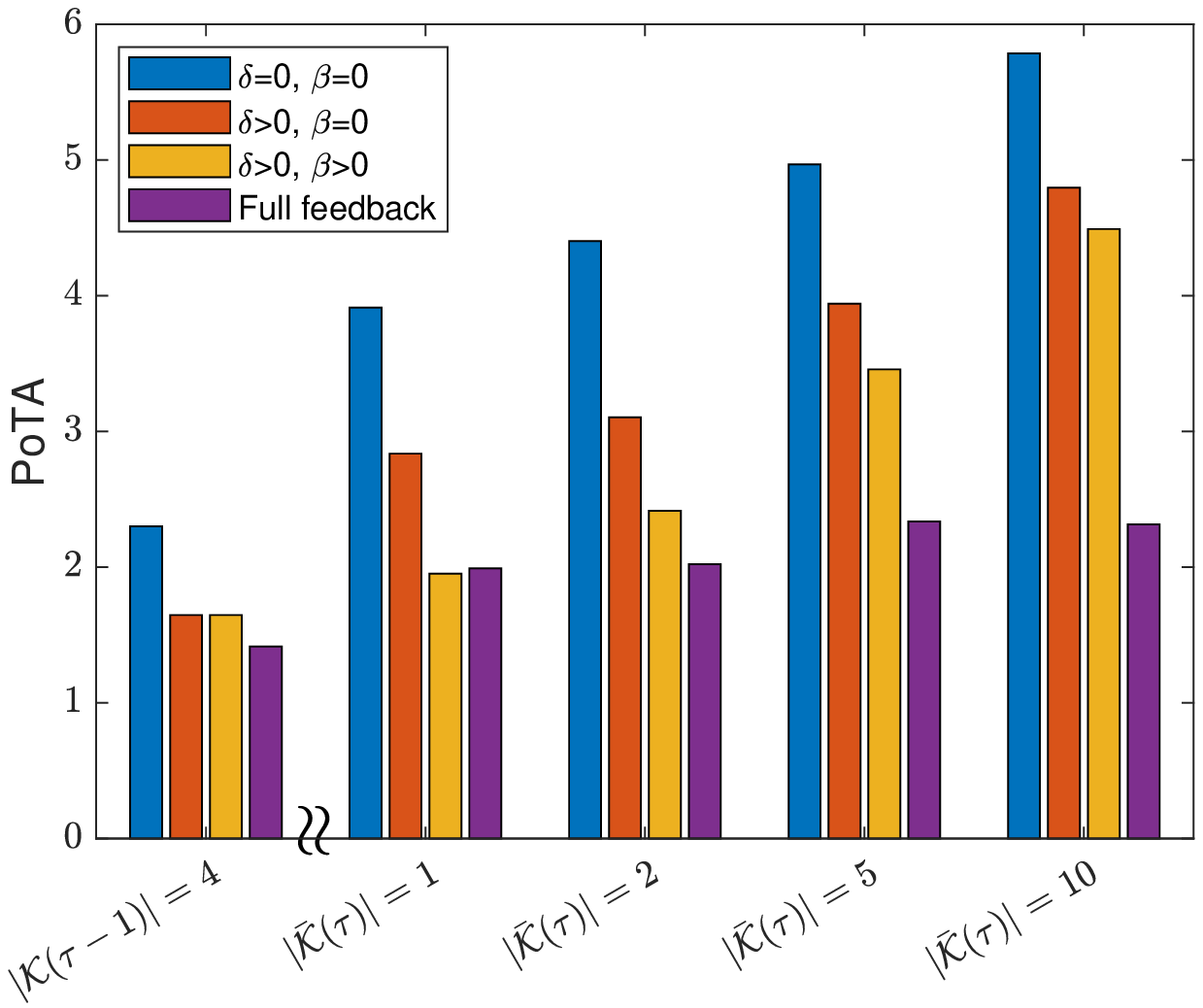}} & \hspace{-20pt} 
					\subfigure[\tiny \label{numerical:demand:benchmark}]
					{\includegraphics[width=0.265\textwidth]{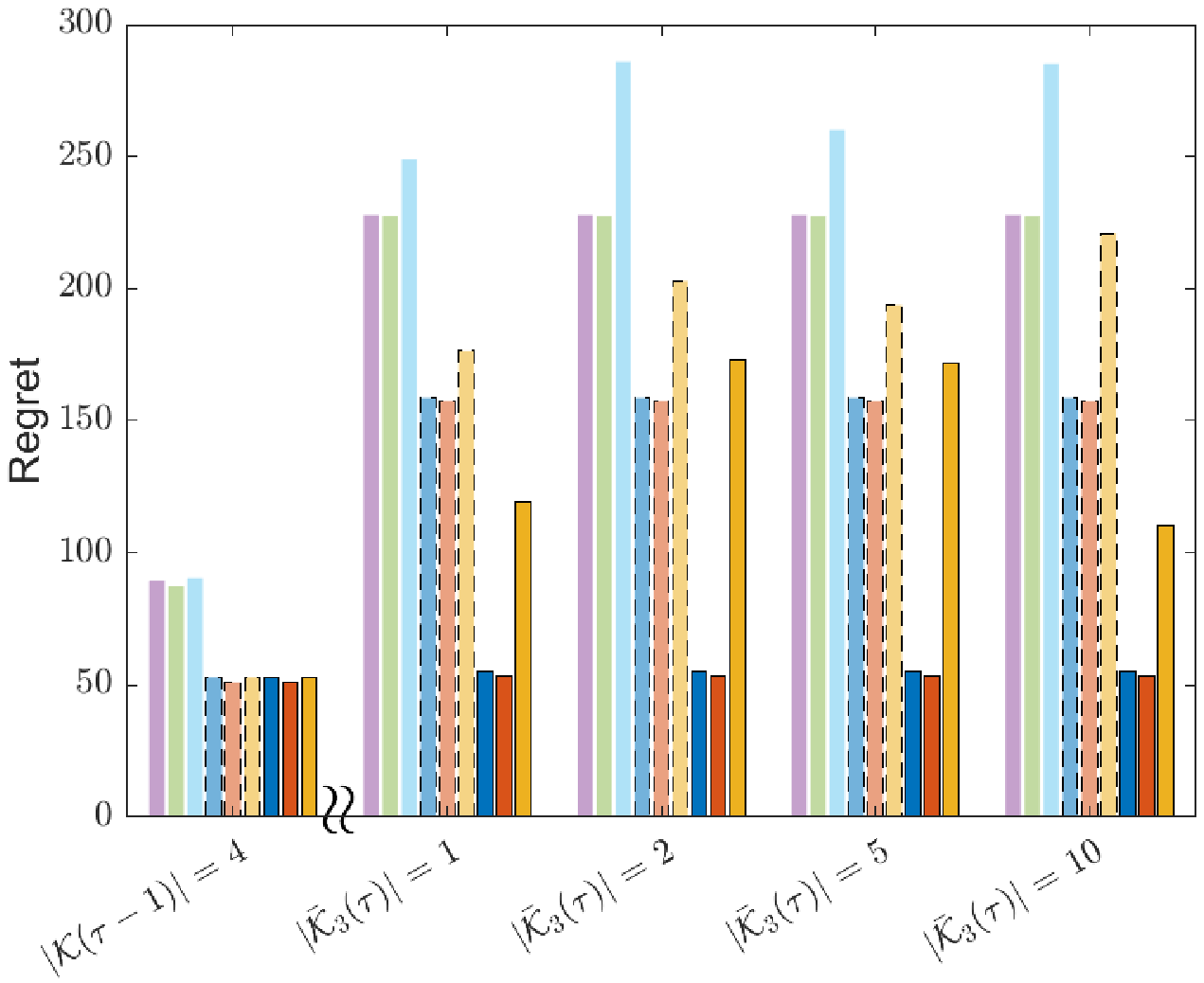}} &
					\hspace{-20pt}
					\subfigure[\tiny \label{numerical:demand:regret}]
					{\includegraphics[width=0.265\textwidth]{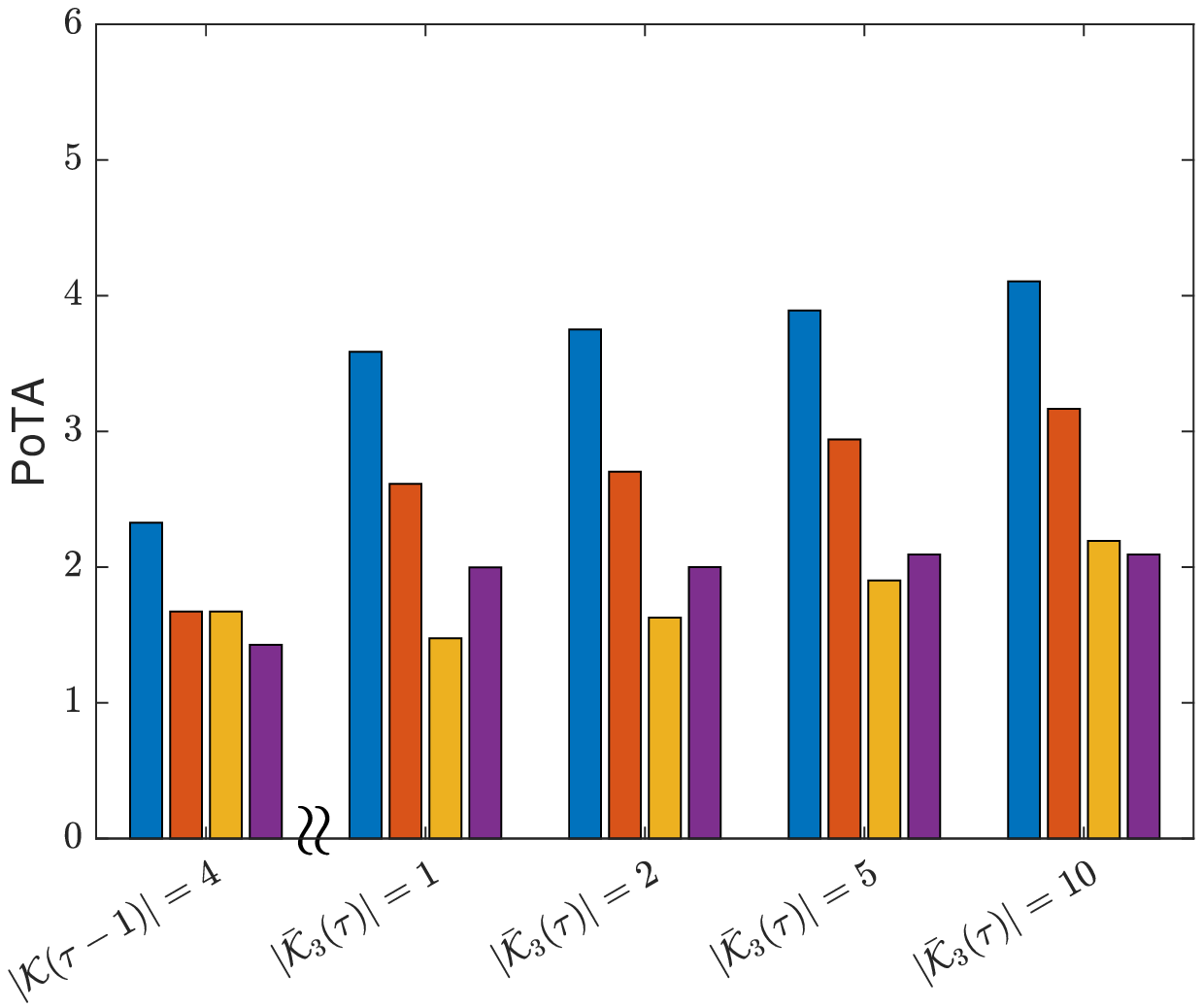}} 
				\end{tabular}\vspace{-1em}
				
				\caption{Impact of arms on regret and PoTA. (Upper) symmetric arms $\mathcal{K}_1({\tau}) = \mathcal{K}_2({\tau})= \mathcal{K}_3({\tau})$ and (Bottom) asymmetric arms $\mathcal{K}_1({\tau}) = \mathcal{K}_2({\tau})\neq \mathcal{K}_3({\tau})$, while $\mathcal{K}_1({\tau-1}) = \mathcal{K}_2({\tau-1}) = \mathcal{K}_3({\tau-1})$.} \vspace{-1em}
				\label{numerical:asymmetric_beta}
			\end{figure*}

			\subsubsection*{{Impact of $\beta$ and $\mathcal{K}$}}
			Fig.~\ref{numerical:asymmetric_beta} shows the impact of the number of VFNs, $|\mathcal{\bar{K}}_n({\tau})|$, appeared in $\tau$, on regret and PoTA. As the density of the candidate VFNs becomes higher, more exploration would be performed, requiring more rounds to make the unit offloading cost converged and resulting in a higher regret for each client. The perturbed exploration taking into account dynamic resource supply, where a client sets the score of a new or re-discovered VFN to its recent update, e.g., the lastly updated score the rejoining VFN had before or the others have, could achieve lower learning regret compared to vanilla Exp3IX with initializing the learning history of all candidates (full-reset case in  \cite{Cho2021}), $\mathcal{\hat{L}}_{nk}({\tau}) = 0, \forall k\in \mathcal{{K}}_n({\tau})$. This is because such a dynamic resource supply-based policy allows for avoiding unfair selection opportunities, i.e., reducing the exploration rounds the appearing arms may require to experience, and thus adapting quickly to the change in a volatile environment. The PoTA enhancement by the perturbed exploration with $\beta>0$ compared to the full reset case with $\beta=0$ decreases in the number of VFNs. The effect of the score difference among the existing VFNs becomes minimal for a high density of the appearing VFNs, since an importance-weighted mechanism assigns a probability proportional to the number of the candidate VFNs as well as the cumulative scores. Note that every task requester may have a different set of VFN candidates due to its inherent attributes such as communication range, mobility, availability, and so on. When only a minority of the clients have increasing candidate VFNs, the system-level performance variation becomes minimal, since another majority of the clients use a dominant strategy.

			\section{Conclusion}
			This work considered the decentralized task offloading decision-making problems of multi-agents in unknown and dynamic environments as a repeated unknown game where each agent has access only to local information and makes an adaptive offloading decision to individually heterogeneous and dynamic situations. Particularly, every agent can take exploration implicitly controlling the variance at the expense of introducing some bias, and observe the current contexts in terms of personalized resource demand and volatile resource supply before playing an action. This work showed that the dynamic behaviors of distributed agents with individual perturbations for robustness to uncertainty (implicit exploration) and adaptation to dynamicity (demand/supply exploration) could converge towards a sequence of stable equilibria and such self-interested decisions ensure better optimality in terms of social welfare, e.g., lowering the upper bounds of PoTA. The effectiveness of the proposed algorithm was verified by simulation results. The future effort could be directed at studying joint computation and communication resource selections in real clock time rather than on the interpolated time interval.
			
			\appendices
			\section*{appendix}
			\subsection{Proof of Lemma~\ref{replicator}}\label{ref:replicator}
			We show that the derivative of the continuous-time limit of the algorithm is the replicator equation. We consider the effect of client $n$'s action $p_{nk}(\tau)$ on its own probability update on $k\in \mathcal{K}_n(\tau)$. We obtain the continuous time process from the rate of change of $p_{nk}(\tau)$ w.r.t $\kappa(\tau)\eta_n(\tau)$ as $\kappa(\tau)\eta_n(\tau) \rightarrow 0$. 	
			
			i) The selection probability is expressed: when $k = k'$ and $m \in \mathcal{K}\backslash k'$, 
				${p}_{nk}(\tau+1) = \frac{e^{-\mathcal{W}_{nk}(\tau)}}{  e^{-\mathcal{W}_{nk}(\tau)}+\sum_{m\in \mathcal{K}\backslash k} e^{-\mathcal{W}_{nm}(\tau)}}  = \frac{e^{-\mathcal{W}_{nk}(\tau-1) }A_{nk}(\tau)}{  e^{-\mathcal{W}_{nk}(\tau-1) }A_{nk}(\tau)+\sum_{m\in \mathcal{K}\backslash k} e^{-\mathcal{W}_{nm}(\tau-1)}} {=}\frac{A_{nk}(\tau)}{  A_{nk}(\tau)- 1+ 1/{p}_{nk}(\tau) } \nonumber
				$ where $A_{nk}(\tau) = e^{-   \zeta_n \kappa(\tau) \eta_n(\tau) \hat{l}_{nk}(\tau)}$, and when $k \in \mathcal{K}\backslash k'$ and $m =  k'$, 			
				${p}_{nk}(\tau+1)	= \frac{e^{-\mathcal{W}_{nk}(\tau)}}{  e^{-\mathcal{W}_{nm}(\tau)}+\sum_{m'\in \mathcal{K}\backslash m} e^{-\mathcal{W}_{nm'}(\tau)}}\\			
				{=} \frac{e^{-\mathcal{W}_{nk}(\tau-1) } }{  e^{-\mathcal{W}_{nm}(\tau-1) }A_{nm}(\tau)- e^{-\mathcal{W}_{nm}(\tau-1) } +\sum_{m'\in \mathcal{K}} e^{-\mathcal{W}_{nm'}(\tau-1)}} \\
				=\frac{p_{nk}(\tau) }{ p_{nm}(\tau) A_{nm}(\tau)- p_{nm}(\tau) + 1}$.			
			ii) The expected update in the probabilities is the following differential equation, $\dot{p}_{nk}$: when $k' = k$, 	
			
			\begin{array}{ll}
				\dot{p}_{nk}  & =   \frac{\partial {p}_{nk}(\tau+1)}{\partial \kappa(\tau)} = \frac{\partial}{\partial \kappa(\tau)} \left( \frac{A_{nk}(\tau)}{1 /{p_{nk}(\tau)- 1+ A_{nk}(\tau)}}\right), \nonumber\\
				& = \frac{\frac{\partial A_{nk}(\tau)}{\partial \kappa(\tau)}   ({1 /{p}_{nk}(\tau)- 1+ A_{nk}(\tau)}) - A_{nk}(\tau) \frac{\partial A_{nk}(\tau)}{\partial \kappa(\tau)} }{({1 /{p}_{nk}(\tau) - 1+ A_{nk}(\tau)})^2},\nonumber\\
				&	= \frac{\partial A_{nk}(\tau)}{\partial \kappa(\tau)}   \frac{ ({1 /{p}_{nk}(\tau)- 1 }) }{({1 /{p}_{nk}(\tau) - 1+ A_{nk}(\tau)})^2},\nonumber 
			\end{array} 
			
			and	when $k' = m \neq k$	 
			
			\begin{array}{ll} 
				\dot{p}_{nk} &\!\!\!\!=    \frac{\partial}{\partial \kappa(\tau)} \left(\frac{{p}_{nk}(\tau)}{1 + {p}_{nm}(\tau) (A_{nm}(\tau)-1)}\right)  \nonumber\\
				&\!\!\!\!=  \frac{\partial A_{nm}(\tau)}{\partial \kappa(\tau)} \frac{-   {p}_{nk}(\tau) {p}_{nm}(\tau) }{(1 + {p}_{nm}(\tau) (A_{nm}(\tau)-1))^2}, \nonumber 
			\end{array} 
		
			where $	\frac{\partial A_{nk}(\tau)}{\partial \kappa(\tau)}     = \frac{-    \zeta_n  \eta_n(\tau) {l}_{nk}(\tau) p_{nk}(\tau)}{(p_{nk}(\tau) + \kappa(\tau) \gamma_n(\tau))^2} e^{-   \kappa(\tau) \eta_n(\tau) \hat{l}_{nk}(\tau)}$.
			
			iii) The continuous time process is obtained by taking the limit ${\kappa(\tau) \rightarrow 0}$, i.e., the rate of change in $p_{nk}$ with respect to ${\kappa(\tau)}$ as ${\kappa(\tau) \rightarrow 0}$. Then, with dropping the discrete time index script $\tau$, we derive the limit of the probability update rule as follows: a) when $k' = k$, $\lim_{\kappa(\tau) \rightarrow 0}	\dot{p}_{nk}   
			= \frac{-    \zeta_n  \eta_n {l}_{nk}}{p_{nk}}  \frac{ ({1 /{p}_{nk}- 1 }) }{({1 /{p}_{nk}})^2} = {-   \zeta_n  \eta_n {l}_{nk}}(1-{p_{nk}})$ and b) when $k' = m \neq k$, $\lim_{\kappa(\tau) \rightarrow 0}	\dot{P}_{nk} = \frac{  \zeta_n   \eta_n {l}_{nm}}{p_{nm}}   {p}_{nk}  {p}_{nm} =   \zeta_n   \eta_n  {l}_{nm}  {p}_{nk}$.   
			
			iv) The expected change in $p_{nk}$ w.r.t the probability distribution of a client over all VFNs $k\in \mathcal{K}_n$ is given as
			
			\begin{array}{ll}
				\!\!\!\!\!\!	E_n[\dot{{p}}_{nk}] \!=\! E_n\left[ 
				\sum_{m \neq k\in \mathcal{K}_n}\! \zeta_n  \eta_n  {l}_{nm}    {p}_{nk}
				-    \zeta_n  \eta_n  {l}_{nk} (1 - {p}_{nk})\right],  \nonumber\\
				=  -    \zeta_n {p}_{nk} \eta_n  {l}_{nk} (1 - {p}_{nk}) + 
				\sum_{m \neq k\in \mathcal{K}_n}   \zeta_n {p}_{nm} \eta_n  {l}_{nm}    {p}_{nk},
				\nonumber\\
				=  \zeta_n \eta_n  {p}_{nk} \left[ \sum_{m \neq k\in \mathcal{K}_n}     {l}_{nm}{p}_{nm}      -   \sum_{m \neq k\in \mathcal{K}_n}     {l}_{nk} {p}_{nm} \right],				\nonumber\\
				=  \zeta_n \eta_n  {p}_{nk} \left[ \sum_{m  \in \mathcal{K}_n}    {l}_{nm}{p}_{nm}  -  {l}_{nk}\right],\nonumber\\
				=  \zeta_n \eta_n  {p}_{nk} \left[ \sum_{m  \in \mathcal{K}}    {l}_{nm}{p}_{nm}  -  {l}_{nk}\right].
			\end{array}

			\subsection{Proof of Lemma~\ref{apt}}\label{ref:apt}
			The actual recursive forms of the algorithm are of the stochastic approximation type. The dynamic of Algorithm 1 can be written as follows:
			Using Taylor's Remainder Theorem, 
			
			\begin{array}{ll} 
				\!\!\!\!\!\!\!\!	p_{nk}(\tau+1) \!=\! \Lambda_{nk}(\mathcal{W}_n(\tau))
				\!=\!  \Lambda_{nk}(\mathcal{W}_n(\tau\!-\!1) \!+\!  \zeta_n \kappa(\tau)\eta_{n}(\tau)\hat{l}_n(\tau)),\nonumber\\
				=   p_{nk}(\tau) +      \kappa(\tau)\left[   \eta_{n}(\tau) \nabla \Lambda_{nk}^T (\mathcal{W}_n(\tau-1)) {l}_n(\tau) \right.\nonumber\\
				~~~~~~~~~+         \eta_{n}(\tau)  \nabla  \Lambda_{nk}^T (\mathcal{W}_n(\tau-1)) (\hat{l}_n(\tau) - {l}_n(\tau)) \nonumber\\	
				~~~~~~~~~\left.+ 
				0.5   {\kappa(\tau)\eta_{n}^2(\tau)} {\hat{l}_n^T(\tau)} \nabla^2 \Lambda_{nk} (\mathcal{W}_n(\tau-1)) {\hat{l}_n(\tau)}\right],\nonumber\\
				= p_{nk}(\tau) + \kappa(\tau)\left[F_{nk}(\tau)  +  \sigma_{nk}(\tau)\right], \nonumber
			\end{array}
			where $F_{n}(\tau)$ is the mean replicator dynamic (Lemma \ref{replicator}), 
			$F_{nk}(\tau)  
			\!=\! \eta_n(\tau) \nabla \Lambda_{nk}^T(\mathcal{W}_n(\tau-1))  	 {l}_{n}(\tau) 
			\!=\!  \eta_n(\tau)  {\zeta_n} \Lambda_{nk}(\mathcal{W}_n(\tau-1)) ( \sum_{m}  \Lambda_{nm}(\mathcal{W}_n(\tau-1)) {l}_{nm}(\tau) - {l}_{nk}(\tau))$ 
			due to ${\partial \Lambda_{nk} (\mathcal{W}_n)}  
			\!=\!  \zeta_n\Lambda_{nk} (\mathcal{W}_n) ( \Lambda_{nm}(\mathcal{W}_n) - \mathbbm{1}_{k=m})$.

			i) (Lipschitz condition) The derivative of the logit choice map function, $\nabla\Lambda_{nk} \left({\zeta_n}  (\mathcal{\hat{L}}_{n}+ \beta_{nk})\right)$, is continuously differentiable with respect to $\mathcal{\hat{L}}_{n}$, which is locally bounded so its gradient is locally bounded as well.
			By changing the input of the derivative function by some amount $\mathcal{\hat{L}}_{n}$, its output changes by at most a multiple of $\mathcal{\hat{L}}_{n}$, known as the Lipschitz constant, a measure of the smoothness of the derivative function. The logit choice map function is the gradient of the log-sum-exp function, $lse(\hat{\mathcal{L}}_{n}) = -\zeta_n^{-1} \log(\sum_k \exp(-\zeta_n \hat{\mathcal{L}}_{nk}))$, $\Lambda_n( {\mathcal{W}}_n) = \nabla lse(\hat{\mathcal{L}}_n)$, its convexity is well-known \cite[p.93]{Boyd2004}. The Jacobian of the logit choice map function is the Hessian of the log-sum-exp function, $\nabla\Lambda_n( {\mathcal{W}}_n) = \nabla^2 lse(\hat{\mathcal{L}}_n)$. 	
			According to \cite[p.58]{Nesterov2004}, a function has a Lipschitz continuous gradient with Lipschitz constant $L>0$ if $0 \geq v^T  \nabla^2 lse(z) v \geq -L ||v||_2^2$ is bounded for all $z, v$, which is fulfilled by the following relation: $ v^T  \nabla^2 lse(z) v = 	-\zeta_n (\sum_k v_k^2 \Lambda_{nk}(z) - (\sum_k v_k\Lambda_{nk}(z))^2)		
			\geq 	-\zeta_n \sum_k v_k^2 \Lambda_{nk}(z) \geq 	-\zeta_n \sum_k v_k^2$ due to {negative semidefinite}, thus  $|v^T  \nabla^2 lse(z) v| \leq |\zeta_n| ||v||_2^2$ and $\nabla\Lambda_{n}(\cdot)$ is a Lipschitz continuous gradient with $\zeta_n$.			
			
			ii) (noise condition)	The term $\sigma_{nk}(\tau)$ is the noise perturbation, $\sigma_{nk}(\tau) =    \eta_n(\tau) \nabla \Lambda_{nk}^T(\mathcal{W}_n(\tau-1))  	  (\hat{l}_n(\tau) - l_n(\tau)) + \frac{\kappa(\tau)\eta_{n}^2(\tau)}{2} \hat{l}_n(\tau) \nabla^2 \Lambda_{nk}\hat{l}_n(\tau)$. The noise term $\sigma_(\tau)$ admits the decomposition, for $i\geq 0$, $\sigma_{n}(\tau) = \chi_{n}(\tau) + \epsilon_{n}(\tau)$
			where $\chi_{n}(\tau)$ is a mean sequence and $\epsilon_{n}(\tau)$ is a bias. If $\chi_{n}(\tau)$ is averaged out by step-sizes $\kappa(\tau)\eta_n(\tau)\geq 0$, $\lim_{\tau\rightarrow\infty} \max_{\tau\leq j < a(\tau,t)} \left| \sum_{z=\tau}^j \kappa(z)\eta_{n}(z) \chi_{n}(z) \right| = 0$, and if $\epsilon_{n}(\tau)$ is bounded, $
			\lim_{\tau\rightarrow\infty}\sup \left| \epsilon_{n}(\tau)\right| < \infty$, then the limits of slightly perturbed solutions to the ODE are an invariant set, called chain recurrent. The sequence $\chi_{nk}(\tau)$ is a martingale difference noise, $\chi_{nk}(\tau) =  \eta_n(\tau) \nabla \Lambda_{nk}^T(\mathcal{W}_n(\tau-1))  	  (\bar{l}_n(\tau) - l_n(\tau))$ due to the following relation, $E[ \chi_{nk}(\tau)|\mathcal{F}_{n,\tau-1}] = E[\eta_n(\tau) \nabla \Lambda_{nk}^T(\mathcal{W}_n(\tau-1))  	  (\bar{l}_n(\tau) - l_n(\tau))|\mathcal{F}_{n,\tau-1}] =  \eta_n(\tau) \nabla \Lambda_{nk}^T(\mathcal{W}_n(\tau-1))  	  ( {l}_n(\tau) - l_n(\tau)) = 0$, which fulfills the zero-mean sequence conditioned on past information, $\mathcal{F}_{n,\tau-1}$. 
			The sequence $\epsilon_{n}(\tau)$ can be further divided into two sub errors, $\epsilon_{n}(\tau) = \epsilon_{n}^1(\tau) + \epsilon_{n}^2(\tau)$ where the error $\epsilon_{n}^1(\tau)$ comes from an implicit bias estimation, $\epsilon_{nk}^1(\tau) =  \eta_n(\tau) \nabla \Lambda_{nk}^T(\mathcal{W}_n(\tau-1))  	  (\hat{l}_n(\tau) - \bar{l}_n(\tau)) < \infty$	
			and the error $\epsilon_{nk}^2(\tau)$ comes from a stochastic approximation, $	\epsilon_{nk}^2(\tau) =  
			\frac{\kappa(\tau)\eta_{n}^2(\tau)}{2} \hat{l}_n(\tau)\nabla^2  \Lambda_{nk}\hat{l}_n(\tau)<\infty$
			due to the facts that all components of $\nabla \Lambda_{nk}^T$ in   $\epsilon_{nk}^1(\tau)$ and $\nabla^2 \Lambda_{nk}^T$ in $\epsilon_{nk}^2(\tau)$ are bounded \cite{Cohen2017}, by definition of the considered implicit bandit estimator. Due to $\frac{\partial \epsilon_{nk}^1}{\partial p_{nk}} <0$, we have $\epsilon_{nk}^1 = \lim_{i\rightarrow\infty} \sup |\eta_n(\tau)|$. And with $\kappa(\tau) \rightarrow 0$, $\epsilon_{nk}^2 \rightarrow 0$ due to bounded components having a type of ${c'}/({p_{nk}(\tau) + \kappa(\tau) \gamma_{n}(\tau)})$ where $0<c'<\infty$. Thus, there exists a non-decreasing function $\phi$ such that $\lim_{\tau\rightarrow\infty} d(p_n(\tau), \mathcal{R}) \leq \phi(\epsilon)$ where $\epsilon$ is a bounded error that tends to an asymptotic bias, $\epsilon = \lim_{\tau\rightarrow\infty} \sup |\epsilon_n(\tau)|<\infty$. Then, according to \cite[Theorem 2.1]{Tadic}, the iterative process $p_n(\tau)$ converges to the internally chain recurrent set of the mean-field system, within a vicinity to $\mathcal{R}$. 
			
			\subsection{Proof of Lemma~\ref{stable}}\label{ref:stable}
			A fixed point of a system is an attractor, when the cost function $ {l}(\Lambda(\mathcal{W}))$ is $||\cdot||_{\infty}$-contractive, a contraction
			map of a locally compact metric space, that is, there exists a constant\footnote{The iterative update is a contracting iteration, if every agent satisfies the condition individually \cite[Lemma 1]{Cho2016}, without knowing the number of other agents. Intuitively, the contractive condition depends on the two different Lipschitzness of the logit function $\Lambda_n$ and the game's cost vector $l_n$. Also, if an extended system with $\mathcal{K}_n = \mathcal{K}, \forall n\in\mathcal{N}$ is contracting, the reduced system $\mathcal{K}_n \subseteq \mathcal{K}, \forall n\in\mathcal{N}$ is also contracting \cite{Cho2016}.} $\hat{\omega}\in(0,1)$ such that for all agents $n\in\mathcal{N}$ and all score variables $\hat{\mathcal{L}}$, $\hat{\mathcal{L}}' \in \mathbb{R}^{|\mathcal{N}|\times|\mathcal{K}|}$, $|| {l}(\Lambda(\mathcal{W}))- {l}(\Lambda(\mathcal{W}'))||_{\infty} \leq \hat{\omega} ||\hat{\mathcal{L}} - \hat{\mathcal{L}}'||_{\infty}$. Consider a cost function, $\theta$-Lipschitz in the congestion degree $c_k$, i.e., a function of a number of agents using $k$-th VFN, $c_k = \sum_{n\in\mathcal{N}} \mathbbm{1}(k = k_{n})$ where $k_n\in\mathcal{K}_n$ w.r.t $||\cdot||_1$ norm: $|l_{nk}(c_k) - l_{nk}(c_k')|   = |l_{n}(k;k_{-n}(c_k)) - l_{n}(k; k_{-n}(c_k'))| \leq \theta ||c_k-c_k'||_1$ for all pairs $k_{-n}(c_k)$, $k_{-n}(c_k') \in \mathcal{K}_{-n}$. According to \cite{Cominetti2010}, the two difference costs for different scores can be expressed as follows: 	
			$|l_{nk}(\mathcal{W}) - l_{nk}(\mathcal{W}')| \leq \theta \sum_{u\neq n} || \Lambda_{u}(\mathcal{W}_u)  - \Lambda_{u}(\mathcal{W}_u') ||_1 \leq \theta\sum_{u\neq n} \sum_{m\in \mathcal{K}_u}||\nabla \Lambda_{um}||_1 \Theta_u \leq  
			\theta\sum_{u\neq n} \sum_{m\in \mathcal{K}_u}
			\left(\zeta_n\Lambda_{nm}   \sum_{k\in \mathcal{K}_n}| \Lambda_{nk}  - \mathbbm{1}_{m=k}|\right)  \Theta_u\leq  
			\theta \sum_{u\neq n} \sum_{m\in \mathcal{K}_u}
			\left(2\zeta_n\Lambda_{nm}   (1- \Lambda_{nm})\right)  \Theta_u \leq  
			\theta \sum_{u\neq n} \sum_{m\in \mathcal{K}_u}
			2\zeta_n \Lambda_{nm}  \Theta_u\leq  
			\theta \sum_{u\neq n} 2\zeta_n ||\hat{\mathcal{L}}_u - \hat{\mathcal{L}}_u'||_{\infty}\leq  
			2\theta \zeta ||\hat{\mathcal{L}} - \hat{\mathcal{L}}'||_{\infty}$ 
			where $\Theta_u = ||\hat{\mathcal{L}}_u - \hat{\mathcal{L}}_u'||_{\infty}$, $\zeta = max_n \zeta_n$ and $\theta$ is an upper bound for the impact over a client's cost when a single client changes its move for each client $n \in \mathcal{N}$.  	
			According to \cite[Theorems 6.9 and 6.10]{Benaim1999} and \cite[Theorem 4]{Cominetti2010}, the converged limit set is contained in every attractor under the logit rule, asymptotically stable for its ODE.
			
			\subsection{Proof of Corollary~\ref{corollary1}}\label{ref:corollary1}
			In case that the cost function is linear in $c_k$, e.g., $l_{nk}  = l_{nk}^o+ \theta \cdot c_k$ where $l_{nk}^o$ is communication cost and $\theta \cdot c_k$ is computation cost upper-bounded by the worst service capability and the number of agents choosing VFN $k$, we have the following relation: $\sum_{m\in \mathcal{K}_u}||\nabla \Lambda_{um}||_1   =  \sum_{m\in \mathcal{K}_u}\zeta_n\Lambda_{nm}   \sum_k| \Lambda_{nk}  - \mathbbm{1}_{m=k}| = \sum_{m\in \mathcal{K}_u}2\zeta_n\Lambda_{nm} (1-\Lambda_{nm}) \leq 2\zeta_n\Lambda_{nk}(1-\Lambda_{nk})\leq \zeta_n/2$ where the last inequality is due to $x(1-x)\leq 1/4, \forall x\in [0,1]$, and thus, $|l_{nk}(\mathcal{W}) - l_{nk}(\mathcal{W}')|  \leq \theta \sum_{u\neq n} \sum_{m\in \mathcal{K}_u} ||\nabla \Lambda_{um}||_1 ||\hat{\mathcal{L}}_u - \hat{\mathcal{L}}_u '||_{\infty} \leq \theta \sum_{u\neq n}   \zeta_u/2   ||\hat{\mathcal{L}}_u - \hat{\mathcal{L}}_u'||_{\infty} \leq \theta \zeta/2  ||\hat{\mathcal{L}} - \hat{\mathcal{L}}'||_{\infty}$. 
			
			\subsection{Proof of Proposition~\ref{prop1_convergeNE}}\label{ref:prop1_convergeNE}
			(NE) We show by contradiction that $\lim_{T\rightarrow \infty}  {p}(T) =  {p}'$ is a NE. Suppose the contrary, there exists $\Delta_{l}>0$ such that $l_{nm} ({p}') > l_{nk} ({p}'), 
			\forall m\in \mbox{supp}(p_n'), 
			k\in \mathcal{K}_n, k\notin \mbox{supp}(p_n')$. The support of $p'$ is the set of pure strategies that have a positive probability of being selected, 
			denoted by $supp(p_n') = \{k_n\in \mathcal{K}_n: p_n>0\}, \forall n\in \mathcal{N}$.	Thus, 
			
			\begin{array}{llll}
				p_{nm}(T+1) = \Lambda_{nm}(\mathcal{ {W}}_{n}(T)) = {e^{-\mathcal{ {W}}_{nm}(T)}}/{\sum_{k} e^{- {\mathcal{W}}_{nk}(T)}} \\
				\overset{ }{\leq} {e^{ {\mathcal{W}}_{nk}(T) - {\mathcal{W}}_{nm}(T)}} \overset{}{\leq}   e^{ - \zeta_n(T) \cdot \Delta_{l} \sum_{\tau=1}^{T} \kappa(\tau)\eta_n(\tau) + \zeta_n(T) \cdot \Delta_{\beta}},
			\end{array}
			where the last inequality relation is due to the following:
			
			\begin{array}{llll}
				{\mathcal{W}}_{nk}(T) - {\mathcal{W}}_{nm}(T),\\ 
				= \zeta_n(T)\sum_{{\tau}=1}^T [\kappa(\tau)\eta_n(\tau)  (\hat{l}_{nk}(\tau)-\hat{l}_{nm}(\tau))] + \zeta_n(T)\cdot \Delta_{\beta}, \\
				\leq  \zeta_n(T)   \sum_{{\tau}=1}^T [\kappa(\tau)\eta_n(\tau)   (-\Delta_{l}) ] + \zeta_n(T) \cdot \Delta_{\beta},
			\end{array}
			where $\Delta_{\beta} = \beta_{nk}(0)-\beta_{nm}(0)$, if i) $\sum_{\tau} \kappa(\tau)\eta_n(\tau) \rightarrow \infty$, ii) $\kappa(\tau)\gamma_n(\tau) > {\tau}^{-1}$, and iii) ${l}_{nm}(\tau) - {l}_{nk}(\tau) \geq \Delta_{l}$ according to \cite[Prop.3]{Cho2021}, Lemma \ref{replicator} and Lemma \ref{apt}. Then, $\lim_{{T}\rightarrow \infty} p_{nm}(T) = 0$ which is a contradiction with $m\in \mbox{supp}(p_n')$, and ${p}'$ is a NE. Thus, the actual sequence $p(\tau)$ converges toward a NE of $\Gamma$. 
			
			($\xi$-equilibrium) According to Definition \ref{logit-equilibrium}, the NE can be satisfied with approximated one with margin $\xi$. Similar to the NE, all agents need not change its strategy profile, in $\xi$-equilibrium. According to \cite{McKelvey1995}, for any $\zeta_n>0$ and $p_n$, the logit choice map $\Lambda(\cdot)$ is $\xi$-approximate with $\xi = \max_{n\in\mathcal{N}}(\log(|\mathcal{K}_n|)/\zeta_n)$, if the score value of the option it picks is at most the minimum value plus $\xi$. In other words, agents do not change their strategy profiles when they cannot obtain more than $\xi$ from other deviations, i.e., an option of value larger than the minimum plus $\xi$ is never chosen.
			
			\subsection{Proof of Proposition~\ref{prop:convergencerate}}\label{ref:prop:convergencerate}	
			$\lim_{T\rightarrow \infty} \boldsymbol{p}(T)$ is a pure state of the form $\boldsymbol{p}^*(T) = \{\Lambda_{nk'}(T)\}_{n\in \mathcal{N}}$ expressed as follows:     
			\begin{array}{llll}
				\Lambda_{nk'}(T)
				= \frac{e^{-\mathcal{W}_{nk'}(T)}}{\sum_k e^{-\mathcal{W}_{nk}(T)}}
				= {1}/({\sum_k e^{\mathcal{W}_{nk'}(T)-\mathcal{W}_{nk}(T)}}),\\
				=   {1}/{({1+\sum_{k \neq k'} e^{\mathcal{W}_{nk'}(T)-\mathcal{W}_{nk}(T)}})},\\
				\geq   {1}/{({1+ (K-1)\max_{k \neq k'} e^{\mathcal{W}_{nk'}(T)-\mathcal{W}_{nk}(T)}})},\\
				\overset{(i)}{\geq} {1}/{({1+ (K-1) e^{-\zeta_n\cdot \{\beta_{nk}(0) - \beta_{nk'}(0) + \Delta_{l}\cdot \sum_{{\tau}=1}^T \kappa(\tau)\eta_n(\tau) \}}})},\\
				\overset{(ii)}{\geq} {1} - (K-1) e^{-\zeta_n\cdot \{\beta_{nk}(0) - \beta_{nk'}(0) + \Delta_{l}\cdot \sum_{{\tau}=1}^T \kappa(\tau)\eta_n(\tau)\}},\\
			\end{array}
			where (i) is due to the sub-relations: ${\mathcal{W}}_{nk'}(T) - {\mathcal{W}}_{nk}(T) \leq    -\zeta_n\cdot \{\beta_{nk}(0) - \beta_{nk'}(0) + \Delta_{l}\cdot \sum_{{\tau}=1}^T \kappa(\tau)\eta_n(\tau) \}$ and  ${l}_{nk}(\tau)- {l}_{nk'}(\tau) \geq \Delta_{l}>0$, and (ii) is due to the relation $1/(1+x)\geq 1-x, x>0$.  Then $k'$ is a.s a strict equilibrium of $\Gamma$ and convergence occurs at a quasi-exponential rate.  
			
			\subsection{Proof of Proposition~\ref{async}}\label{ref:async}			
			
			Note that the clients inactive at ${\tau}$ do not update. We assume that $\lim_{{\tau}\rightarrow\infty} \mbox{inf}~ \vartheta_n(\tau)/{\tau} >0, \forall n \in \mathcal{N}$ satisfied if $\mathcal{N}(\tau)$ is an irreducible Markov chain over $2^{\mathcal{N}}:= \{\mathcal{N}'\subseteq\mathcal{N}\}$. Thus, when ${\tau}\rightarrow\infty$, we have $\vartheta_n(\tau) \rightarrow \infty$, i.e., $\sum_{{\tau}\geq 0}^{\infty} \mathbbm{1}_{n\in\mathcal{N}(\tau)} = \infty$. 	
			For any $n\in\mathcal{N}$, if $\sum_{\tau}\kappa(\tau)\eta_n(\tau) = 	\infty$, we have
			$ \sum_{{\tau}}\kappa(\tau)\eta_n(\tau) = \sum_{\tau} 
			\kappa(\vartheta_n(\tau))\eta_n(\vartheta_n(\tau))\mathbbm{1}_{n\in\mathcal{N}(\tau)}	\leq  \sum_{{\tau}} [\max_{n\in\mathcal{N}}\kappa(\vartheta_n(\tau))\eta_n(\vartheta_n(\tau))] = \sum_{\tau} \kappa^*(\tau) = \infty$. 
			We have $\sum_{{\tau}} [\kappa^*(\tau)]^2 = \sum_{{\tau}} [\max_{n\in \mathcal{N}(\tau)} \kappa(\vartheta_n(\tau))\eta_n(\vartheta_n(\tau))]^2 
			\leq  \sum_{{\tau}}\sum_{n\in\mathcal{N}} [\kappa(\vartheta_n(\tau))\eta_n(\vartheta_n(\tau))]^2  \mathbbm{1}_{n\in\mathcal{N}(\tau)} \leq |\mathcal{N}|\sum_{\tau} \kappa^2(\tau)\eta_n^2(\tau)<\infty$, if $\sum_{\tau} \kappa^2(\tau)\eta_n^2(\tau)<\infty$.

			\subsection{Proof of Proposition~\ref{pota_bound}}\label{ref:pota_bound}				
			Consider a sequence of strategies generated by repeated play. For any smooth game of individual clients $n\in\mathcal{N}$ for stage ${\tau}\in\mathcal{T}$, one may have the following inequalities: 			
			\begin{array}{llll}
				\!{\sum_{{\tau}}C({\tau})} \!\! = {\sum_{{\tau}}\sum_{n}} l_{nk'}(\tau) 
				= {\sum_{{\tau}}\sum_{n}} l_{n}(k'(\tau);\Theta), \\
				= {\sum_{{\tau}}\sum_{n}}  [l_n({k}^{*}; \Theta) ]  + \sum_{\tau}\sum_{n}  [l_{nk'}(\tau) \!-\! l_n({k}^{*}; \Theta)], \\ 
				\leq \sum_{{\tau}} [\lambda(\tau)   C^* + \mu(\tau)   C(\tau)]  + \sum_{\tau}\sum_{n}  [l_{nk'}(\tau) \!-\! l_n({k}^{*}; \Theta)], \\ 
				\leq \sum_{{\tau}} [\lambda(\tau)   C^* + \mu(\tau)   C(\tau)]  + \sum_{\tau}\sum_{n}  [l_{nk'}(\tau) \!-\! l_n({k}''; \Theta)], \\ 
				\leq \sum_{{\tau}} [\lambda'   C^* + \mu'   C(\tau)]  + \sum_{\tau}\sum_{n}  [l_{nk'}(\tau) \!-\! l_n({k}''; \Theta)], \\ 
				\leq \frac{{\lambda'{\sum_{{\tau}}}  C^*}}{1-\mu'}  \!+\! \frac{{\sum_{n}}\sum_{{\tau}}}{(1-\mu')}   [l_{nk'}(\tau) \!-\! l_n({k}''; \Theta)],		\\	
				= \frac{{\lambda'{\sum_{{\tau}}}  C^*}}{1-\mu'}  \!+\! \frac{{\sum_{n}}}{(1-\mu')}  \sum_{{\tau}} [l_{nk'}(\tau) \!-\! l_{nk''}(\tau)], 
			\end{array}
			where $\Theta = k_{-n}'(\tau)$ and $ l_{nk'}(\tau) = l_{n}(k'(\tau);\Theta)$, a function of $(k'({\tau}); k_{-n}'({\tau}))$, e.g., the cost observed by a client $n$ for ${\tau}$ with bandit feedback, $\sum_{{\tau}}l_{nk''}( \tau) \!\!\!= \sum_{{\tau}}l_n({k}''; {k}_{-n}'({\tau})) \!\!\!= \!\!\! \min_{k} \sum_{{\tau}}l_n({k}; {k}_{-n}({\tau}))$, i.e., the minimum cost a client $n$ could have achieved by playing the best-fixed action in case the sequence $\{k_{-n}({\tau})\}_{{\forall\tau}}$ of others' actions and the cost function were known in hindsight, and $\lambda'$ and $\mu'$ are the worst case smoothness of $\Gamma$, not necessarily context-dependent, e.g., regardless of resource demand and supply \cite{Roughgarden2015}.

			\vspace{1em}
			\bibliographystyle{IEEEtranTCOM}

\begin{thebibliography}{94}
				\setstretch{1.2} 
				
				\bibitem{3gpp2018}
				3GPP, ``Technical Specification Group Services and System Aspects; Enhancement of 3GPP Support for V2X Scenarios,''
				Tech. Rep. TS 22.186, Rel. 15, Sept. 2018.
				
				\bibitem{Mao2017}
				Y. Mao \emph{et~al.}, ``A survey on mobile edge computing: The communication perspective,'' \emph{IEEE Commun. Surveys Tuts.}, vo. 19, no. 4, pp. 2322–2358, Fourth quarter 2017.
				
				\bibitem{Mouradian2018}
				C. Mouradian \emph{et~al.}, ``A comprehensive survey on fog computing: State-of-the-art and research challenges,'' \emph{IEEE Commun. Surveys Tuts.}, vo. 20, no. 1, pp. 416–464, First quarter 2018.
				
				\bibitem{Hou2016}
				X. Hou \emph{et~al.}, ``Vehicular Fog Computing: A Viewpoint of Vehicles as the 
				Infrastructures,'' \emph{IEEE Trans. Veh. Technol.}, vo. 65, no. 6, pp. 3860-3873, 2016. 
				
				\bibitem{Zhu2018}
				C. Zhu \emph{et~al.}, ``Vehicular Fog Computing for Video Crowdsourcing: Applications, Feasibility, and Challenges,'' \emph{IEEE Comm. Mag.}, vo.56, no.10, pp. 58-63, 2018. 
				
				\bibitem{Chen2019}
				T. Chen \emph{et~al.}, ``Learning and management for Internet of Things: Accounting for adaptivity and scalability,'' \emph{Proc.~the IEEE}, vo. 107, no. 4, pp. 778-796, 2019.
				
				\bibitem{Cho2021}
				B. Cho \emph{et~al.}, ``Learning-based decentralized offloading decision making in an adversarial environment,'' \emph{IEEE Trans. Veh. Technol.}, vo. 70, no. 11, pp. 11308-11323, 2021. 
    
				\bibitem{Zheng2015}
				K. Zheng \emph{et~al.}, ``An SMDP-based resource allocation in vehicular cloud computing systems,'' \emph{IEEE Trans. Ind. Electron.}, vo. 62, no. 12, pp. 7920–7928, 2015. 
				
				\bibitem{Gu2018}
				B. Gu \emph{et~al.}, ``A distributed and context-aware task assignment mechanism for collaborative mobile edge computing,'' \emph{Sensors}, vo. 18, no. 8, 2018. 
				
				\bibitem{Zhou2021a}
				S. Zhou \emph{et~al.}, ``Machine learning-based offloading strategy for lightweight user mobile edge computing tasks,'' \emph{Complexity}, 2021.
				
				\bibitem{Jehangiri2022}
				S. Zaman \emph{et~al.}, ``LiMPO: lightweight mobility prediction and offloading framework using machine learning for mobile edge computing,'' \emph{Cluster Computing}, pp. 1–19, 2022.
				
				\bibitem{Feng2018}
				J. Feng \emph{et~al.}, ``Mobile edge computing for the internet of vehicles: Offloading framework and job scheduling,'' \emph{IEEE Veh. Technol. Mag.}, vo. 14, no. 1, pp. 28-36, 2018.
				
				\bibitem{sun2019}
				Y. Sun \emph{et~al.}, ``Adaptive learning based task offloading for vehicular edge computing systems,'' \emph{IEEE Trans. Veh. Technol.}, vo. 68, no. 4, pp. 3061–3074, 2019.
				
				\bibitem{Zhu2019} 
				Z. Zhu \emph{et~al.}, ``BLOT: Bandit learning based offloading of tasks in fog-enabled networks,'' \emph{IEEE Trans. Parallel Distrib. Syst.}, vo. 30, no. 12, pp. 2636-2649, 2019.
				
				\bibitem{Liu2020}
				Z. Liu \emph{et~al.}, ``Learning based fluctuation-aware computation offloading for vehicular edge computing system,'' \emph{Proc. IEEE WCNC}, pp. 1-7, 2020.
				
				\bibitem{Zhang2020}
				R. Zhang \emph{et~al.}, ``Online learning enabled task offloading for vehicular edge computing,'' \emph{IEEE Wireless Commun. Lett.}, , vo. 9, no. 7, pp. 928-932, 2020.
				 
				\bibitem{Xiao2020}
				X. Xu \emph{et~al.}, ``Distributed no-regret learning in multiagent systems: Challenges and recent developments,''\emph{IEEE Sig. Proc. Mag.}, vo. 37, no. 3, pp. 84-91,  2020.
				
				\bibitem{sun2021}
				Z. Sun \emph{et~al.}, ``Applications of Game Theory in Vehicular Networks: A Survey,'' \emph{IEEE Commun. Surveys Tuts.}, vo. 23, no. 4, pp. 2660-2710, Fourth quarter 2021.
				 
				\bibitem{Chen2016}
				X. Chen \emph{et~al.}, ``Efficient multi-user computation offloading for mobile-edge cloud computing,''\emph{IEEE/ACM Trans. Netw.}, vo. 24, no. 5, pp. 2795–2808, 2016.	
				
				\bibitem{Zheng2018}
				J. Zheng \emph{et~al.}, ``Dynamic computation offloading for mobile cloud computing: a stochastic game-theoretic approach,''\emph{IEEE Trans. Mob. Comput.}, vo. 18, no. 4, pp. 771–786,  2018.	
				
				\bibitem{Josilo2019}
				S. Jošilo \emph{et.al}, ``Wireless and computing resource allocation for selfish computation offloading in edge computing,''\emph{IEEE INFOCOM}, pp. 2467-2475, 2019.
				  
				\bibitem{Sergiu2000}
				H. Sergiu \emph{et~al.}, ``A simple adaptive procedure leading to correlated equilibrium,'' in \emph{Econometrica}, vo. 68, no 5, pp. 1127-1150, 2000.
				
				\bibitem{Kleinberg2009}
				R. Kleinberg \emph{et~al.}, ``Multiplicative Updates Outperform Generic No-Regret Learning in Congestion Games,'' \emph{Proc. ACM Theory of computing}, pp. 533-542. 2009.
				 
				\bibitem{Krichene2015} 
				W. Krichene \emph{et~al.}, ``Online learning of Nash equilibria in congestion games'', \emph{SIAM J. Control Optim.}, vo. 53, no. 2, pp. 1056–1081, 2015.
				
				\bibitem{Palaiopanos2017} 
				G. Palaiopanos \emph{et~al.}, ``Multiplicative weights update with constant step-size in congestion games: Convergence, limit cycles and chaos'', \emph{Proc. Neural Information Processing Systems}, 2017.
				
				\bibitem{Coucheney2015}
				P. Coucheney \emph{et~al.}, ``Penalty-regulated dynamics and robust
				learning procedures in games'', \emph{Mathematics of Operations Research}, vo. 40, no. 3, pp. 611–633, 2015.
				
				
				\bibitem{Cohen2017}
				J. Cohen \emph{et~al.}, ``Learning with bandit feedback in potential games'', \emph{Proc. Neural Information Processing Systems}, pp. 6372-6381, 2017.
				 
				\bibitem{Feng2017}
				X. Feng \emph{et.al}, ``Distributed cell selection in heterogeneous wireless networks,''  \emph{Computer Communications}, 109, pp.13-23, 2017.
				
				\bibitem{Milchtaich1996}
				I. Milchtaich, ``Congestion games with agent-specific payoff functions'', \emph{Games Econ. Behav.}, vo. 13, no. 1, pp. 111–124, 1996. 

				\bibitem{Gummadi2013}
				R. Gummadi \emph{et~al.}, ``Mean ﬁeld analysis of multi-armed bandit games'', available at SSRN 2045842, 2013.
				 
				\bibitem{Shi2021}
				C. Shi \emph{et.al}, ``On No-Sensing Adversarial Multi-agent Multi-armed Bandits with Collision Communications''  \emph{IEEE J. Sel. Areas Information Theory}, vo. 2, no. 2, pp. 515-533, 2021.
				 
				\bibitem{Zhou2021}
				Z. Zhou \emph{et.al}, ``Learning-based URLLC-aware task offloading for internet of health things'', \emph{IEEE J. Sel. Areas Commun.}, vo. 39, no. 2, pp. 396-410, 2020.
				 
				\bibitem{2016Akherfi} 
				K. Akherfi \emph{et~al.}, ``Mobile cloud 	computing for computation offloading: Issues and challenges,'' \emph{Applied Computing \& Informatics}, 2016.
				 
				\bibitem{2018Neto}
				J. L. D. Neto \emph{et~al.}, ``ULOOF: A user level online offloading framework for mobile edge computing,'' \emph{IEEE Trans. Mobile Computing,} vo. 17, no. 11, pp. 2660-2674, 2018.
				 
				\bibitem{cesa2006}
				N. Cesa-Bianchi \emph{et~al.}, \emph{Prediction, learning, and games}, {Cambridge university press}, 2006.
				 
				\bibitem{Kenney2011}
				J. B. Kenney \emph{et~al.}, ``Dedicated short-range communications standards in the united states,'' \emph{Proc. the IEEE}, vo. 99, no. 7, pp. 1162-1182, 2011.
				
				\bibitem{Rai2007}
				V. Rai \emph{et~al.}, ``Cross-channel interference test results: A report from the vsc-a project,'' \emph{IEEE 802.11 WG submission 11-07-2133-00-000p}, 2007. 
				
				\bibitem{Ackermann2009}
				H. Ackermann, \emph{Nash Equilibria and Improvement Dynamics in Congestion Games}, Ph.D. dissertation, Universitätsbibliothek, 2009. 
				 
				\bibitem{Glicksberg1952}	
				I. L. Glicksberg, ``A further generalization of the Kakutani fixed point theorem, with application to Nash equilibrium points,''  \emph{Proc. the American Mathematical Society}, vo 3, no. 1, pp. 170-174, 1952.
				 
				\bibitem{McKelvey1995}
				R. D. McKelvey \emph{et.al}, ``Quantal response equilibria for normal form games,''  \emph{Games and economic behavior}, vo. 10, no. 1, pp.6-38, 1995.
				 
				\bibitem{Rosenski2016}
				J. Rosenski \emph{et~al.}, ``Multi-agent bandits–a musical chairs approach,'' \emph{Int. Machine Learning}, vo. 48, pp.155–163, 2016.
				 	
				\bibitem{Fox1972}
				A. J. Fox, ``{Outliers in time series},'' \emph{the Royal Statistical Society: Series B (Methodological)}, vo. 34, no. 3, pp. 350-363, 1972.
				 
				\bibitem{Neu2015}
				G. Neu \emph{et~al.}, ``Explore no more: Improved high-probability regret bounds for non-stochastic bandit,'' \emph{Advances in Neural Information Processing Systems}, 2015.
				
				\bibitem{Taylor1978}
				P. Taylor,  \emph{et~al.}, ``Evolutionarily stable strategies and game dynamics,'' \emph{Math. Biosci},vo. 40, no. 1-2, pp. 145–156, 1978.
				 
				\bibitem{Benaim1999}
				M. Benaim, ``A dynamical system approach to stochastic approximations,'' \emph{SIAM J. Control Optim.}, vo. 34, no.2, pp.437-472, 1996.
				
				\bibitem{Ljung1977}
				L. Ljung, ``Analysis of recursive stochastic algorithms,'' \emph{IEEE Trans. Automatic Control}, vo. 22, no. 4, pp. 551-575, 1977.
				
				\bibitem{Kushner1977}
				H. Kushner, ``General convergence results for stochastic approximations via weak convergence theory,'' \emph{J. Math. Anal. Appl.}, no. 61, pp. 490–503, 1977.
				
				\bibitem{Kushner1984}
				H. Kushner \emph{et~al.}, ``An invariant measure approach to the convergence of stochastic approximations with state dependent noise,'' \emph{SIAM J. Control Optim.}, no. 22, 13–27, 1984. 
				
				\bibitem{LaSalle1960}
				J.-P. LaSalle, ``Some Extensions of Liapunov’s Second Method,'' \emph{IRE Transactions on Circuit Theory}, vo. 7, pp. 520–527, 1960.
				
				\bibitem{Michel1999}
				Michel. B, \emph{Dynamics of stochastic approximation algorithms}, Séminaire de probabilités de Strasbourg, 33, 1999.
				
				\bibitem{Tadic}
				V. Tadic \emph{et~al.}, ``Asymptotic bias of stochastic gradient search,'' \emph{The Annals of Applied Probability}, vo. 27, no. 6, pp.3255-3304, 2017. 
				
				\bibitem{Boyd2004}
				S. Boyd \emph{et~al.}, \emph{Convex optimization, 1st ed. Cambridge},
				UK: Cambridge University Press, 2004.
				
				\bibitem{Nesterov2004}
				Y. Nesterov, \emph{Introductory Lectures on Convex Optimization: A Basic
					Course}, Norwell, MA: Kluwer, 2004.
				
				\bibitem{Cho2016}
				B. Cho \emph{et~al.}, ``Co-primary spectrum sharing for inter-operator device-to-device communication,'' \emph{IEEE J. Sel. Areas Commun.}, vo. 35, no. 1, pp.91-105, 2016.
				 
				\bibitem{Cominetti2010}
				R. Cominetti \emph{et~al.}, ``A payoff-based learning procedure and its application to trafﬁc games,'' \emph{Games and Economic Behavior}, vo. 70, no. 1, pp. 71-83, 2010.
				
				\bibitem{Kushner2003}
				H. Kushner \emph{et~al.}, \emph{Stochastic approximation and recursive algorithms and applications}, vo. 35., Springer Science \& Business Media, 2003
				 
				\bibitem{Roughgarden2015}	
				T. Roughgarden, ``Intrinsic robustness of the price of anarchy,'' \emph{Journal of the ACM}, vo. 62, no. 5, pp. 1-42, 2015.
				 
				\bibitem{3GPP2011}
				3GPP TR 36.931, ``Evolved universal terrestrial radio access,'' v.9.0.0, \emph{Tech. Rep.}, 2011.
				
			\end{thebibliography}

  \end{document}